%% file: main.tex
\documentclass[reprint,amsmath,amssymb,aps]{revtex4-2}

\usepackage[pdftex]{graphicx}
\usepackage[FIGTOPCAP,nooneline]{subfigure}
\subcapraggedrighttrue
\usepackage{dcolumn}
\usepackage{bm}
\usepackage{braket}
\usepackage[version=3]{mhchem}
\usepackage{hyperref}
\usepackage{xcolor}
\hypersetup{
    colorlinks=true,
    citecolor=blue,
    linkcolor=blue,
    urlcolor=blue,
}
\usepackage{url}
\usepackage{mathrsfs}
\allowdisplaybreaks
\usepackage{multirow}   
\usepackage{float}       
\usepackage{cancel}

\begin{document}

\preprint{APS/123-QED}

\title{Nonlinear Edelstein Effect in Strongly Correlated Electron Systems}
\author{Jun {\=O}ik{\'e}} 
\email{oike.jun.32y@st.kyoto-u.ac.jp}
\author{Robert Peters}
\affiliation{Department of Physics, Kyoto University, Kyoto 606-8502, Japan}

\date{\today}

\begin{abstract}
Nonlinear spintronics, which combines nonlinear dynamics and spintronics, opens a new route for controlling spins and spin dynamics beyond conventional spintronics based on linear responses.
Strongly correlated electron systems, which can have large nonlinear responses, are promising candidates for nonlinear spintronics.
In this paper, we focus on the nonlinear Edelstein effect (NEE), a generalization of the Edelstein effect, and study the impact of electron correlations on the NEE by performing numerical calculations on a Hubbard model.
We find that correlation effects can either enhance or suppress the nonlinear response. We show that the enhancement and suppression of the response are due to the real and imaginary components of the self-energy, respectively. 
Additionally, we have explored the relationship between the NEE and photomagnetic or optomagnetic effects. 
Our findings demonstrate that electron correlations can either enhance or suppress the optical spin injection, depending on the light frequency, while always strengthening the inverse Faraday effect.
\end{abstract}

\maketitle

\input{sections/section01.tex}
\input{sections/section02.tex}
\input{sections/section03.tex}

\input{sections/section04.tex}
\input{sections/section05.tex}
\input{sections/section06.tex}
\section*{Acknowledgments} \label{sec:Acknowledgments}
J.O. is grateful to Koki Shinada and Taisei Kitamura for their valuable comments.
R.P. is supported by JSPS KAKENHI No.~JP23K03300. Parts of the numerical simulations in this work have been done using the facilities of the Supercomputer Center at the Institute for Solid State Physics, the University of Tokyo.

\appendix

\begin{widetext}
\input{appendices/appendixA.tex}
\input{appendices/appendixB.tex}
\input{appendices/appendixC.tex}
\input{appendices/appendixD.tex}

\bibliography{main.bib}

\end{document}

%% file: sections/section01.tex
\section{Introduction} \label{sec:Introduction}

Electric generation and control of spin degrees of freedom are central issues in modern spintronics \cite{Utic2004-za,Bader2010-my,Hirohata2020-mr}. 
A typical technique is to utilize the linear Edelstein effect (LEE) where an electric field $\bm{E}$ induces a nonequilibrium spin density $\delta \bm{s}$ in noncentrosymmetric metals  ($\delta \bm{s} \propto \bm{E}$). 
The LEE was first theoretically proposed in Refs. \cite{A_G_Aronov_and_Yu_B_Lyanda-Geller1989-su,Edelstein1990-qc} and has experimentally been realized in GaAs \cite{Kato2004-rw,Silov2004-tm,Ganichev2006-pd}. 
Furthermore, a spin torque driven by the LEE has theoretically been proposed in magnetic semiconductors \cite{Manchon2008-to,Manchon2009-qh,Garate2009-fn}, 
and a current-induced domain inversion has been observed in GaMnAs \cite{Chernyshov2009-yj}, which has triggered increased attention to the LEE in the branch of spintronics.


Interestingly, the literature concerning the LEE deals almost exclusively with noninteracting and weakly correlated electron systems 
such as topological insulators \cite{Pesin2012-vr,Mellnik2014-oe,Ando2014-eu,Li2014-ke,Rodriguez-Vega2017-cv,Cullen2023-ag}, Weyl semimetals \cite{Johansson2018-ns,Zhao2020-pr,Limtragool2022-dv,Burkov2023-bk}, and superconductors \cite{Edelstein1995-ba,He2020-hz,Ikeda2020-vi,Nakajima2023-mz,Shinada2023-uv,Yuan2023-wz}; 
only a few works have analyzed electron correlations using Fermi liquid theory and dynamical mean-field theory (DMFT) \cite{Fujimoto2007-ht,Fujimoto2007-fl,Yanase2014-sx,Peters2018-mm}. 
One reason is that conventional spintronics has been based on linear responses.
It has been demonstrated in Ref. \cite{Michishita2021-yw} that correlation effects do not strongly affect linear DC conductivities.
On the other hand, large nonlinear conductivities have been realized in strongly correlated electron systems (SCES) in experiments \cite{Kishida2000-zz,Liu2016-wi,Dzsaber2021-uw,Uchida2022-lk,Murakami2022-ne} and numerical calculations \cite{Murakami2018-wr,Tancogne-Dejean2018-ah,Michishita2021-yw,Kofuji2021-ti}. 
Therefore, SCES would be one of the most promising candidates for nonlinear spintronics.


In recent years, nonlinear responses have also been considered in the field of spintronics.
Of particular interest is the nonlinear Edelstein effect (NEE) \cite{Xiao2023-yu,Baek2023-aq,Feng2024-gj,Guo2024-pu,Battiato2014-hx,Berritta2016-fo,Freimuth2016-kq,Xu2021-hx,Fregoso2022-fd},
characterized by $\delta \bm{s} \propto \bm{E}^2$.  
The NEE significantly influences the spin response from the following three points. 
First, the NEE has no restrictions on spatial inversion $\mathcal{P}$. 
Since the spin is an axial vector but the electric field is a polar vector, the LEE can only appear in noncentrosymmetric materials. 
On the other hand, the NEE exists in centrosymmetric systems, which account for approximately 80\% of all materials. 
Second, the NEE can also be observed in semiconductors and insulators \cite{Xu2021-hx,Fregoso2022-fd}. 
Optical transitions can generate a static spin density in the same way as the bulk photovoltaic effect generates a static electric polarization \cite{Koch1976-ra,Von_Baltz1981-ah,Dai2023-ft}.
Thus, the NEE can be divided into the current-induced NEE \cite{Xiao2023-yu,Baek2023-aq,Feng2024-gj,Guo2024-pu}, which can be observed in metals similarly to the LEE, and the light-induced NEE \cite{Battiato2014-hx,Berritta2016-fo,Freimuth2016-kq,Xu2021-hx,Fregoso2022-fd}, which also can be finite in semiconductors and insulators. 
Note that we here define the static response in nonmagnetic materials as the Edelstein effect and the static response in magnets as the magnetoelectric effect \cite{Curie1894-yo,Xiao2022-xr} regardless of whether they are metals or insulators.
Third, the NEE can have a sizable magnitude despite its nonlinear nature. With a driving electric field of $E=10^5\sim 10^7\ \mathrm{V/m}$ which is readily feasible in an experiment \cite{Chernyshov2009-yj}, the strength of the NEE would exceed that of the LEE in transition metal dichalcogenides \cite{Xu2021-hx} and common transition metals \cite{Xiao2023-yu,Baek2023-aq}.


The advantages of SCES for nonlinear responses combined with the advantages of the NEE over the LEE have the potential to make the NEE in SCES a new important research target in spintronics.
In this work, 
we study the impact of electron correlations on the NEE by performing numerical calculations on a Hubbard model. 
We first formulate the NEE at finite temperatures based on a diagrammatic approach \cite{Parker2019-pa,Joao2019-fb,Michishita2021-yw,Du2021-sc,Wang2022-jy}. 
This formalism allows us to derive equations using single-particle Green's functions whose self-energy includes correlation effects. 
We then use DMFT \cite{Georges1996-cq} to obtain the self-energy, and analyze the response functions of the NEE incorporating electron correlations. 
Note that we do not consider the orbital contribution by itinerant electrons, i.e., the orbital Edelstein effect \cite{Yoda2015-yq,Yoda2018-pr,Niu2019-un,He2020-ch,Hara2020-pb} 
because there is an inherent difficulty in defining the orbital angular momentum operator in periodic crystals \cite{Thonhauser2005-wp,Xiao2005-bu,Shi2007-xs}, 
and the spin contribution usually dominates the total magnetization \cite{Meyer1961-tf,reck1969orbital}.


The rest of this paper is organized as follows: In Sec. \ref{sec:Formulation}, we introduce the formalism of the NEE to the second order in an electric field. 
Section \ref{sec:Model and Method} presents the details of the Hubbard model, which we use to analyze the NEE, and shows spectral functions of the model obtained by DMFT.
We then show numerical results of the interaction dependence for the current-induced NEE and the light-induced NEE in Sec. \ref{sec:results}. 
The results show that the effect of electron correlations on the NEE depends on the form of the applied electric field. Specifically, correlation effects can enhance the current-induced NEE but can either enhance or suppress the light-induced NEE.
Section. \ref{sec:Discussion} is devoted to discussing the results of the previous section.
In Secs. \ref{subsec:Discussion-A} and \ref{subsec:Discussion-B}, we reveal that the enhancement and suppression originate from the real and imaginary parts of the self-energy.
Furthermore, we explain how the real-part and imaginary-part effects contribute to the light-induced NEE and to what degree they affect it.
In Sec. \ref{subsec:Discussion-C}, we remark on the relationship between the light-induced NEE and photomagnetic or optomagnetic effects studied in the field of magneto-optics.
We here focus on the optical spin injection \cite{Oestreich2002-ax,Utic2004-za} for the photomagnetic effects and the inverse Faraday effect (IFE) \cite{Kimel2005-ir,Kirilyuk2010-ys} for the optomagnetic effects and 
demonstrate that electron correlations can either enhance or suppress the optical spin injection depending on the light frequency while strengthening the IFE regardless of it.
Finally, we summarize this work and give a future outlook in Sec. \ref{sec:Summary and future outlook}.

%% file: sections/section02.tex
\section{Formulation} \label{sec:Formulation}

The spin density induced by an electric field is written in the frequency domain as 
\begin{align}
    &\braket{\delta\hat{s}_\alpha(\omega_\Sigma)}=\int \frac{d \omega_1}{2\pi} \zeta^{(1)}_{\alpha;\beta}(\omega_\Sigma;\omega_1)E^\beta(\omega_1)2\pi\delta_{\omega_1,\omega_\Sigma} \notag \\
    &+\int  \frac{d\omega_1}{2\pi} \frac{d\omega_2}{2\pi} \zeta^{(2)}_{\alpha;\beta \gamma }(\omega_\Sigma;\omega_1,\omega_2) E^\beta(\omega_1) E^\gamma(\omega_2) 2\pi \delta_{\omega_1+\omega_2,\omega_\Sigma} \notag \\
    &+\cdots , \label{SPIE}
\end{align}
where $\zeta^{(n)}_{\alpha;\alpha_1 \cdots \alpha_n }(\omega_\Sigma;\omega_1,\cdots,\omega_n)$ is the $n$-th order response function, and Greek indices label Cartesian components. 
$\omega_\Sigma$ corresponds to the frequency of the generated spin response, and $\omega_i$ corresponds to the frequency of the electric field, $E$.
In this study, we focus on the static response ($\omega_\Sigma=0$) recognized as the Edelstein effect.
The first term on the right side of Eq. \eqref{SPIE} represents the LEE, while all other terms describe the NEE.
In particular, the second-order NEE can be divided into two cases:
\begin{align}
    \zeta^{(2)}_{\alpha;\beta\gamma}&=\lim_{\omega_1,\omega_2 \to 0 }\zeta^{(2)}_{\alpha;\beta\gamma}(\omega_\Sigma;\omega_1,\omega_2), \label{CNEE} \\
    \zeta^{(2)}_{\alpha;\beta\gamma}(\Omega)&=\zeta^{(2)}_{\alpha;\beta\gamma}(0;\Omega,-\Omega),\label{LNEE}
\end{align}
where $\Omega$ is the frequency of the incident light. 
Equations \eqref{CNEE} and \eqref{LNEE} describe the current-induced NEE and the light-induced NEE, respectively. 
We derive Eqs. \eqref{CNEE} and \eqref{LNEE} based on the path integral Matsubara formalism as shown below.


The unperturbed Hamiltonian of the system is 
\begin{align}
    \label{General Hamiltonian}
    \hat{H}(\bm{k})=\hat{H}_0(\bm{k})+\hat{H}_{\mathrm{int}},
\end{align}
where $\hat{H}_0(\bm{k})$ is a noninteracting Hamiltonian, and $\hat{H}_{\mathrm{int}}$ is the two-particle interacting part of the Hamiltonian. 
Throughout this paper, we suppose that there is only a local interaction that does not depend on the momentum $\bm{k}$.


Next, we consider the interaction between carriers and the electromagnetic field. 
We here assume the electric dipole approximation under which the electromagnetic field is approximated by a uniform electric field $\bm{E}(t)$.  
This uniformity limits the gauge degree of freedom to either the length or velocity gauge.
We here adopt the velocity gauge where we can treat the effect of the electric field by rewriting Eq. (\ref{General Hamiltonian}) as 
\begin{align}
    \label{minimal coupling}
    \hat{H}(\bm{k}) &\rightarrow \hat{H}(\bm{k}-\frac{q}{\hbar}\bm{A}(t)) \notag \\
    &=\hat{H}(\bm{k})+\sum^{\infty}_{n=1} \frac{1}{n!} \Bigl[ \prod^{n}_{i=1} (-\frac{q}{\hbar} A^{\alpha_i}(t) \partial_{k_{\alpha_i}}) \Bigr] \hat{H}(\bm{k}),
\end{align}
where $\bm{E}(t)=-\partial\bm{A}(t) /\partial t$, and $q$ is the charge of the carriers. 
We expand the powers of the vector potential in a Taylor series to capture nonlinear responses in the second line of Eq. (\ref{minimal coupling}). 


Furthermore, we include an auxiliary term $\hat{H}_B=-\bm{B}(t)\cdot \hat{\bm{s}}$ in Eq. \eqref{General Hamiltonian} to obtain the spin response.
$\bm{B}(t)$ is the conjugate field of the spin $\hat{\bm{s}}$ and is taken to zero after the variation. 
Note that one could derive similar formulations for other physical quantities by replacing $\hat{\bm{s}}$ and $\bm{B}(t)$ with the quantity of interest $\hat{\bm{\theta}}$ and its conjugate field, respectively.
However, $\hat{\bm{\theta}}$ must be a local operator well-defined in periodic systems such as the spin operator $\hat{\bm{s}}$.


The partition function of the perturbed system is written in the path integral formalism as 
\begin{align}
    \label{partition function}
    Z[A,B]=\int \mathcal{D} \bar{\psi} \mathcal{D} \psi\mathrm{exp} \bigl[-S[A,B] \bigr],
\end{align}
where $\bar{\psi}$ and $\psi$ are fermionic creation and annihilation operators, represented by Grassmann numbers, and $S[A,B]$ is the action of the system in imaginary time ($\tau$), which is described as
\begin{align}
    &S[A,B]=\int^{\beta}_0 d\tau \Biggl[ \sum_{\lambda,\eta} \int \frac{d\bm{k}}{(2\pi)^d} \bar{\psi}_{\bm{k} \lambda}(\tau) \Bigl\{ (\partial_{\tau}-\mu)\delta_{\lambda \eta} \notag \\
    &+  H_0^{\lambda \eta}(\bm{k}-\frac{q}{\hbar}\bm{A}(\tau))  -\bm{B}(\tau) \cdot \bm{s}^{\lambda \eta} \Bigr\} \psi_{\bm{k} \eta} (\tau) +H_{\mathrm{int}} \Biggr], 
\end{align} 
where $d$ is the dimension of the system, $\mu$ the chemical potential, and $X^{\lambda \eta} $  the matrix representation of an operator $\hat{X}$.
The expectation value of $\hat{\bm{s}}$ is expressed by the functional derivative \cite{Altland2010-op} as
\begin{align}
    \braket{\delta \hat{s}_{\alpha}(\tau)}=\left. \frac{\delta}{\delta B^{\alpha}(\tau)} \right|_{B=0} \mathrm{ln} Z[A,B].
\end{align}
We then expand $Z[A,B]$ in powers of the vector potential $\bm{A}(\tau)$ and define the $n$-th order response function as the coefficient to $A^{\alpha_1}(\tau_1) \cdots A^{\alpha_n}(\tau_n)$ which is given by 
\begin{align}
    \label{functional derivative}
    &\chi^{(n)}_{\alpha;\alpha_1 \cdots \alpha_n}(\tau;\tau_1 \cdots \tau_n) \notag \\
    &=\left. \frac{1}{Z[0]} \Biggl( \prod^{n}_{i=1} \frac{\delta}{\delta A^{\alpha_i}(\tau_i)}\Biggr) \frac{\delta}{\delta B^{\alpha}(\tau) } \right|_{B=A=0} Z[A,B].
\end{align}
After taking the Fourier transformation to Matsubara frequencies and performing an analytic continuation,
we can obtain the second-order response function expressed by single-particle Green's functions as 
\begin{widetext}
\begin{align}
    \label{Second-order response function}
    &\zeta^{(2)}_{\alpha;\beta \gamma}(\omega_\Sigma;\omega_1,\omega_2) =\frac{\hbar}{(\omega_1+i\delta)(\omega_2+i\delta)} \int \frac{d \bm{k}}{(2\pi)^d} \int^{\infty}_{-\infty} \frac{d \omega}{2\pi i} f(\omega)  \notag \\
    &\times \mathrm{Tr} \biggl[ \frac{1}{2} \Bigl( s_{\alpha} G^R(\bm{k},\omega+\omega_\Sigma) J_{\beta \gamma}(\bm{k})G^{R-A}(\bm{k},\omega)+s_{\alpha} G^{R-A}(\bm{k},\omega) J_{\beta \gamma}(\bm{k}) G^A(\bm{k},\omega-\omega_\Sigma) \Bigr)  \notag \\
    &+s_{\alpha} G^R(\bm{k},\omega+\omega_\Sigma )J_{\beta}(\bm{k}) G^R(\bm{k},\omega+\omega_2) J_{\gamma}(\bm{k})G^{R-A}(\bm{k},\omega)+s_{\alpha} G^R(\bm{k},\omega+\omega_1) J_{\beta}(\bm{k})G^{R-A}(\bm{k},\omega)J_{\gamma}(\bm{k}) G^A(\bm{k},\omega-\omega_2) \notag \\
    &+s_{\alpha} G^{R-A}(\bm{k},\omega) J_{\beta}(\bm{k}) G^A(\bm{k},\omega-\omega_1)J_{\gamma}(\bm{k}) G^A(\bm{k},\omega-\omega_\Sigma)  \biggr] +\bigl[(\beta,\omega_1)\leftrightarrow(\gamma,\omega_2)\bigr] ,
\end{align}
\end{widetext}
where $f(\omega)=(1+\mathrm{exp}(\beta \hbar \omega ))^{-1}$ is the Fermi distribution function, $G^{R/A}(\bm{k},\omega )$ a retarded/advanced Green's function, $J_{\alpha_1 \cdots \alpha_n}(\bm{k} )=(q/\hbar)^{n} \partial_{k_{\alpha_1}} \cdots \partial_{k_{\alpha_n}}H_0(\bm{k})$ a current operator, and $G^{R-A}=G^R-G^A$. 
$\bigl[(\beta,\omega_1)\leftrightarrow(\gamma,\omega_2)\bigr]$ corresponds to an interchange of these indices and frequencies.
Furthermore, we use that the $n$-th order response function is defined as the coefficient relating the spin response to $E^{\alpha_1}(\omega_1) \cdots E^{\alpha_n}(\omega_n)$, which is given by
\begin{align}
    &\zeta^{(n)}_{\alpha;\alpha_1 \cdots \alpha_n}(\omega_\Sigma;\omega_1 \cdots \omega_n) \notag \\
    &=\chi^{(n)}_{\alpha;\alpha_1 \cdots \alpha_n}(\omega_\Sigma;\omega_1 \cdots \omega_n)/\prod^{n}_{j=1} i(\omega_j+i\delta).
\end{align}
because
\begin{align}
    \label{AtoE}
    \bm{E}(\omega_j)=i(\omega_j+i\delta) \bm{A}(\omega_j),
\end{align}
where $\delta>0$ is an adiabatic factor for the external field and is taken to zero after the calculation. 
Note that we ignore vertex corrections in the occurring many-particle Green's function, which enables us to describe the response function as a product of single-particle Green's functions.
The details of the derivation are given in Appendix \ref{app:A}. 
Correlation effects are taken into account through the self-energy $\Sigma^{R/A}(\omega)$ of the Green's function, $G^{R/A}(\bm{k},\omega)=(\hbar \omega-H_0(\bm{k}) +\mu-\Sigma^{R/A}(\omega)+i\eta)^{-1}$,
where $\eta>0$ is an adiabatic factor of the Green's function.
Throughout this paper, we ignore the momentum dependence of the self-energy by using the DMFT approximation \cite{Georges1996-cq}.
Thus, ignoring vertex corrections does not break the generalized Ward identity \cite{Schrieffer1964-op}.
On the other hand, if the self-energy includes momentum dependence, the inclusion of vertex corrections is necessary to satisfy the generalized Ward identity.


We can calculate the current-induced NEE and the light-induced NEE from Eq. \eqref{Second-order response function}. 
For the current-induced NEE, however, it is necessary to correctly analyze the divergence that occurs in the low-frequency region.
If we correctly take the DC limit ($\omega_1,\omega_2 \rightarrow 0$), we can rewrite Eq. \eqref{Second-order response function} as 
\begin{widetext}
\begin{align}
    \label{CNEE_GR}
    \zeta^{(2)}_{\alpha;\beta\gamma} 
    &=-2 \hbar \int \frac{d\bm{k}}{(2\pi)^d} \int^{\infty}_{-\infty} \frac{d \omega }{2\pi } \biggl\{ \biggl(-\frac{\partial f(\omega)}{\partial \omega } \biggr)\mathrm{Im} \biggl( \mathrm{Tr} \biggl[ s_{\alpha} \frac{\partial G^R(\bm{k},\omega)}{\partial \omega } J_{\beta}(\bm{k})G^R(\bm{k},\omega) J_{\gamma}(\bm{k})G^A(\bm{k},\omega) \notag \\
    &+\frac{1}{2} s_{\alpha} \frac{\partial G^R(\bm{k},\omega)}{\partial \omega } J_{\beta\gamma}(\bm{k})G^A(\bm{k},\omega) \biggr] \biggr)-f(\omega)\mathrm{Im} \biggl( \mathrm{Tr} \biggl[ s_{\alpha}\frac{\partial}{\partial \omega} \biggl( \frac{\partial G^R(\bm{k},\omega)}{\partial \omega } J_{\beta}(\bm{k}) G^R(\bm{k},\omega) \biggr)J_{\gamma}(\bm{k}) G^R(\bm{k},\omega) \notag \\
    &+\frac{1}{2} s_{\alpha}\frac{\partial^2 G^R(\bm{k},\omega)}{\partial \omega^2}J_{\beta\gamma}(\bm{k})G^R(\bm{k},\omega )\biggr]\biggr) \biggr\} +(\beta \leftrightarrow \gamma),
\end{align}
\end{widetext}
which is derived in Appendix \ref{app:B}.
We note that for the light-induced NEE, an expression similar to Eq.~\eqref{Second-order response function} has been derived using the Keldysh formalism \cite{Freimuth2016-kq}.
Furthermore, we can reproduce the results of the semiclassical approach \cite{Xiao2022-xr,Xiao2023-yu} and the reduced density matrix formalism \cite{Fregoso2022-fd} by taking the weak-scattering limit in Eqs. \eqref{Second-order response function} and \eqref{CNEE_GR} , which is given in Appendices \ref{app:C} and \ref{app:D}.

%% file: sections/section03.tex
\section{Model and Method} \label{sec:Model and Method}

\begin{figure}[t]
    \setcounter{figure}{1}
    \subfigure[]{
    \includegraphics[height=3.6cm,width=0.425\hsize]{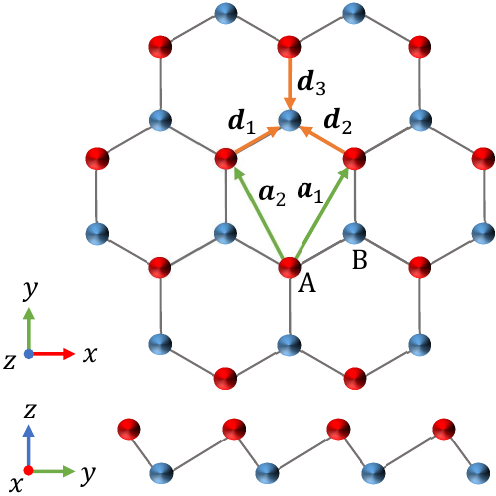}
    \label{Fig:model}
    }
    \subfigure[]{
    \includegraphics[height=3.6cm,width=0.5\hsize]{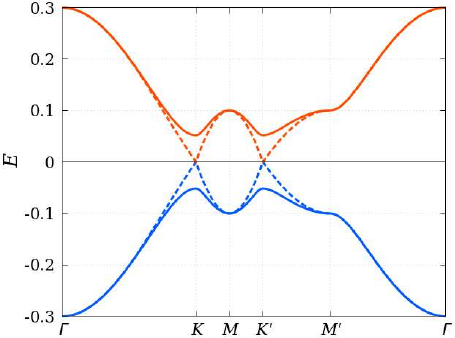}
    \label{Fig:energy_dispersion}
    }
    \setcounter{figure}{0}
    \caption{(a) Top and side views of the model.
                    Green solid arrows show the lattice vectors, and orange solid arrows show the vectors connecting between the NN sites.
             (b) Energy dispersion of the KMH model at $U=0$ for $(t,t_R)=(0.1,0.01)$. The solid lines show the dispersion for $t_{so}=0.1$ and the dashed lines for $t_{so}=0.0$.}
    \label{Fig:model and energy dispersion}
\end{figure}


\subsection{Model} \label{subsec:Model and Method-A}

We here consider a Kane-Mele-Hubbard (KMH) model defined on a buckled two-dimensional honeycomb lattice, which is shown in Fig. \ref{Fig:model}.
The Hamiltonian is
\begin{align}
    \label{model Hamiltonian}
    \hat{H}&=-t\sum_{\braket{ij},\sigma}\hat{c}^{\dag}_{i\sigma} \hat{c}_{j\sigma}+it_{so} \sum_{\braket{\braket{ij}},\sigma\sigma'}\nu_{ij}\hat{c}^{\dag}_{i\sigma} \sigma^{z}_{\sigma \sigma'}\hat{c}_{j\sigma'} \notag \\
    &\hspace{0.4cm} -it_R \sum_{\braket{\braket{ij}},\sigma \sigma'} \mu_{ij} \hat{c}^{\dag}_{i\sigma}(\bm{\sigma}\times\bm{d}_{ij})^z_{\sigma \sigma'} \hat{c}_{j \sigma'}+U \sum_{i} \hat{n}_{i \uparrow} \hat{n}_{i \downarrow}, 
\end{align}
where the noninteracting part is based on a Kane-Mele model \cite{Xiao2023-yu,Kane2005-lh,Liu2011-yq,Liu2011-hs}
and belongs to the $D_{3d}$ point group.
Here, $\hat{c}^{\dag}_{i \sigma}$ and $\hat{c}_{i \sigma}$ are creation and annihilation operators of electrons with spin $\sigma=\uparrow,\downarrow$ at site $i$, $\hat{n}_{i\sigma}=\hat{c}^{\dag}_{i\sigma}\hat{c}_{i\sigma}$ is the number operator, and 
$\sum_{\braket{ij}}$ and $\sum_{\braket{\braket{ij}}}$ are the sum over the nearest neighbor (NN) and next-nearest neighbor (NNN) sites, respectively.
The first term is the NN hopping with hopping strength $t$. The second term is an intrinsic spin-orbit coupling (SOC) with coupling strength $t_{so}$ between NNN electrons, 
where $\sigma^{z}_{\sigma \sigma'}$ is the $z$ component of the Pauli matrix and $\nu_{ij}=+(-)$ if the electron moves counterclockwise (clockwise) around a hexagon.
The third term is the Rashba SOC with coupling strength $t_R$ between NNN electrons, where $\bm{d}_{ij}$ is the unit vector pointing from $j$ to $i$, and $\mu_{ij}=+(-)$ for the A (B) site.
This term originates in the lattice buckling, which causes the symmetry reduction from $D_{6h}$ to $D_{3d}$. A visualization of the lattice is shown in Fig. \ref{Fig:model}.
The last term is a Hubbard-like on-site interaction with interaction strength $U$.


In momentum space, the noninteracting part is given by
\begin{align}
    \label{Bloch Hamiltonian}
    \hat{H}_0&=\sum_{\bm{k},\sigma} ( \eta(\bm{k}) \hat{c}^{\dag}_{\bm{k} A \sigma} \hat{c}_{\bm{k} B \sigma} +\mathrm{h.c.}) \notag \\
    &\hspace{0.4cm}+\sum_{\bm{k},s s,\sigma \sigma'} \bm{g}(\bm{k})\cdot \bm{\sigma}_{\sigma \sigma'} \tau^z_{ss'} \hat{c}^{\dag}_{\bm{k} s \sigma} \hat{c}_{\bm{k} s' \sigma'},
\end{align}
where $\hat{\Psi}_{\bm{k}}=(\hat{c}_{\bm{k} A\uparrow},\hat{c}_{\bm{k} A \downarrow},\hat{c}_{\bm{k} B \uparrow},\hat{c}_{\bm{k} B \downarrow})^t$ is the basis with the momentum $\bm{k}$ and spin $\sigma=\uparrow,\downarrow$ on two sublattices $s=A,B$, and $\bm{\sigma}(\bm{\tau})$ is the Pauli matrix for the spin (sublattice) degrees of freedom. 
The coefficients are defined as
\begin{align}
\eta(\bm{k})&=-t\  \textstyle{\sum^{3}_{i=1}} e^{i\bm{k}\cdot \bm{d}_i}, \label{eta_Hamiltonian} \\
g_x(\bm{k})&=\sqrt{3}t_R (\sin\bm{k}\cdot \bm{a}_1+\sin\bm{k}\cdot \bm{a}_2), \label{gx} \\
g_y(\bm{k})&=-t_R(\sin\bm{k}\cdot \bm{a}_1-\sin\bm{k}\cdot \bm{a}_2 \notag \\
&\hspace{2.0cm} +2\sin\bm{k}\cdot (\bm{a}_1-\bm{a}_2)), \label{gy} \\
g_z(\bm{k})&=2t_{so}(\sin\bm{k}\cdot \bm{a}_1-\sin \bm{k}\cdot \bm{a}_2 \notag \\
&\hspace{2.0cm}-\sin \bm{k}\cdot (\bm{a}_1-\bm{a}_2)),   \label{gz}  
\end{align}
where $\bm{a}_1=(\sqrt{3}a/2,3a/2)$ and $\bm{a}_2=(-\sqrt{3}a/2,3a/2)$ are the lattice vectors,  
$\bm{d}_1=(\sqrt{3}a/2,a/2)$, $\bm{d}_2=(-\sqrt{3}a/2,a/2)$, and $\bm{d}_3=(0,-a)$ are the vectors connecting between the NN sites, and $a$ is the lattice constant.
Equation \eqref{eta_Hamiltonian} is responsible for a linear dispersion at the $K$ and $K'$ points (Dirac points) similar to graphene. 
Equations \eqref{gx}$\sim$\eqref{gz} are the sublattice-dependent antisymmetric SOC in locally noncentrosymmetric systems \cite{Kane2005-lh,Yanase2014-sx,Zhang2014-nc,Zelezny2014-ia,Watanabe2021-kw}
whose site symmetry lacks $\mathcal{P}$ symmetry, while global $\mathcal{P}$ symmetry is preserved by interchanging the sublattice. 
Specifically, Eq. (\ref{gz}) is the SOC that opens a gap at the Dirac points, shown in Fig. \ref{Fig:energy_dispersion}.
Equations (\ref{gx}) and (\ref{gy}) are the Rashba SOC appearing in systems where the site symmetry is denoted by the noncentrosymmetric point group $C_{3v}$,
while global symmetry belongs to the centrosymmetric point group $D_{3d}$.


In numerical calculations, we use the following parameters: $(t,t_{so},t_{R},T)=(0.1,0.01,0.01,0.001)$, where $T$ is the temperature.
Then, we set the Planck constant, the charge of carriers, the Boltzmann constant, and the lattice constant to unity; $\hbar=q=k_B=a=1$.
Furthermore, we use a basis that satisfies the Bloch equation
\begin{align}
    \hat{H}_0(\bm{k}) \ket{u_n (\bm{k})}=\varepsilon_n(\bm{k}) \ket{u_n(\bm{k})},
\end{align}
where $\ket{u_n(\bm{k})}$ is the periodic part of the Bloch state, and $\varepsilon_n(\bm{k})$ is the eigenvalue labeled by the crystal momentum $\bm{k}$ in the first Brillouin zone (BZ).
The index $n=(n,i_n)$ corresponds to a band $n$ and the spinor index $i_n=1,2$.
The matrix representation of an operator $\hat{X}$ under this basis becomes
\begin{align}
    X^{nm}(\bm{k})=\braket{u_{n}(\bm{k})|\hat{X}|u_{m}(\bm{k})}= \bigl[ U(\bm{k})^{-1} X U(\bm{k}) \bigr]^{nm},
\end{align}
where $U(\bm{k})$ is the unitary matrix diagonalizing the noninteracting Hamiltonian $H_0(\bm{k})$, and 
$X$ is the matrix representation under the basis $\hat{\Psi}_{\bm{k}}$.
As a specific example, the spin operator becomes
\begin{align}
    \label{spin_in_Bloch}
    \bm{s}^{nm}(\bm{k})&=\frac{\hbar}{2} \biggl[ \ U({\bm{k}})^{-1} 
    \begin{pmatrix}
        \bm{\sigma} & 0 \\
        0 & \bm{\sigma} \\
    \end{pmatrix}
    U({\bm{k}}) \biggr]^{nm} ,
\end{align}
where $\bm{\sigma}$ is the Pauli matrix for each sublattice.
In the following, we will often omit the $\bm{k}$-index of an operator $X(\bm{k})$ to enhance the readability.


\begin{figure}[t]
    \setcounter{figure}{2}
    \centering
    \subfigure[]{
    \includegraphics[height=3.5cm,width=0.468\hsize]{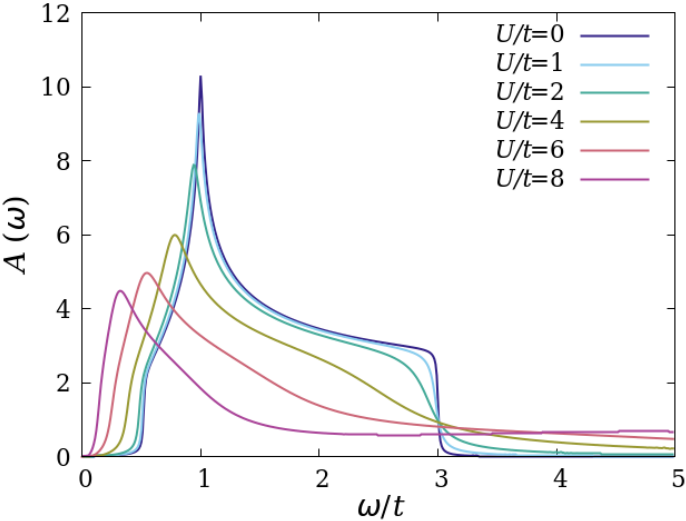}
    \label{fig:spectral_function}
    }
    \subfigure[]{
    \centering\includegraphics[height=3.5cm,width=0.468\hsize]{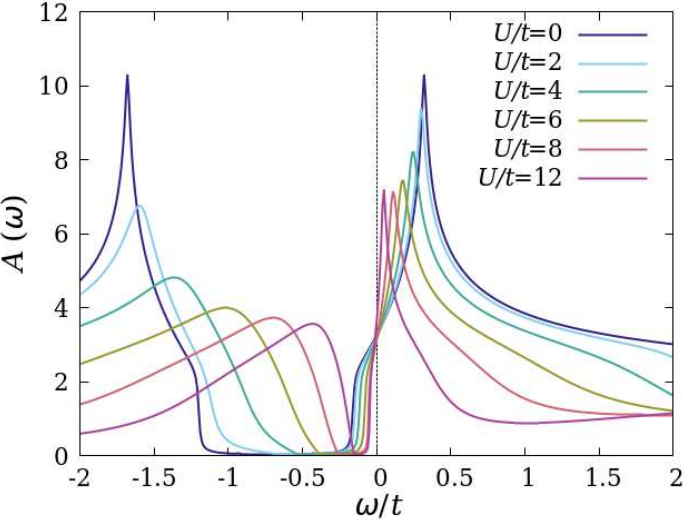}
    \label{fig:spectral_function_dop}
    }
    \setcounter{figure}{1}
    \caption{(a) Spectral functions for different interaction strengths $U/t$ from $0 \sim 8$ at half-filling.  
                Note that only the positive region is shown because of the particle-hole symmetry.
    (b) Spectral functions in the electron-doped regime for different $U/t$ from $0 \sim 12$. We set the electron density to $\langle n\rangle=1.097$.
    For both cases, we use $\eta=0.001$ in Eq. \eqref{Spectral_function}. We note that during the self-consistent cycle, we use larger values to improve the convergence, i.e., (a) $\eta=0.01$ and (b) $\eta=0.02$.
    }
    \label{fig:spectral_function_total}
\end{figure}


\subsection{Dynamical mean-field theory} \label{subsec:Model and Method-B}

DMFT maps the original lattice problem onto a self-consistent quantum impurity problem \cite{Georges1996-cq}.\ 
This mapping is performed by calculating the local Green's function
\begin{align}
    \label{local_Green}
    G(\omega)=\int \frac{d\bm{k}}{(2\pi)^d} \Bigl[\hbar \omega-H_0(\bm{k})+\mu-\Sigma({\omega})+i\eta \Bigr]^{-1}.
\end{align}
In this description, the coupling of the quantum impurity to a bath of conduction electrons, $g(\omega)$, is given as
\begin{align}
\label{self-consistent}
    g^{-1}(\omega)=G^{-1}(\omega)+\Sigma(\omega).
\end{align}
To find the self-consistent solution of the KMH model, we proceed in the following way.
We first prepare a self-energy as an initial value and solve the quantum impurity problem, defined by $g^{-1}(\omega)$ in Eq. \eqref{self-consistent}. 
Then, calculating the local Green's function, Eq.~\eqref{local_Green}, and using Eq.~(\ref{self-consistent}), we obtain an improved $g(\omega)$.
This self-consistent cycle is repeated until convergence is achieved. 
Here, we use the numerical renormalization group \cite{Peters2006-hd,Bulla2008-ok} to solve this quantum impurity problem and to calculate the self-energy.


The spectral function is defined via the Green's function as
\begin{align}
    \label{Spectral_function}
    A(\omega)=-\frac{1}{\pi} \int \frac{d\bm{k}}{(2\pi)^d} \mathrm{Im} G^R(\bm{k},\omega).
\end{align} 
We show spectral functions of the KMH model at half-filling and in the electron-doped regime in Figs. \ref{fig:spectral_function} and \ref{fig:spectral_function_dop}, respectively.
As the interaction increases, the peaks in the spectral function are gradually suppressed, and the spectral weight is continuously transferred to higher energies.
On the other hand, the peak position moves closer to the Fermi energy because of the renormalization of the band structure.
This is the characteristic behavior of correlated electron systems.
In the following, we calculate the current-induced NEE in the electron-doped regime and the light-induced NEE at half-filling.
We note that the current-induced NEE vanishes at half-filling because the system is an insulator.

%% file: sections/section04.tex
\section{results} \label{sec:results}

\begin{figure}[b]
  \setcounter{figure}{3}
  \centering
  \subfigure[]{
  \includegraphics[height=3.2cm,width=0.468\hsize]{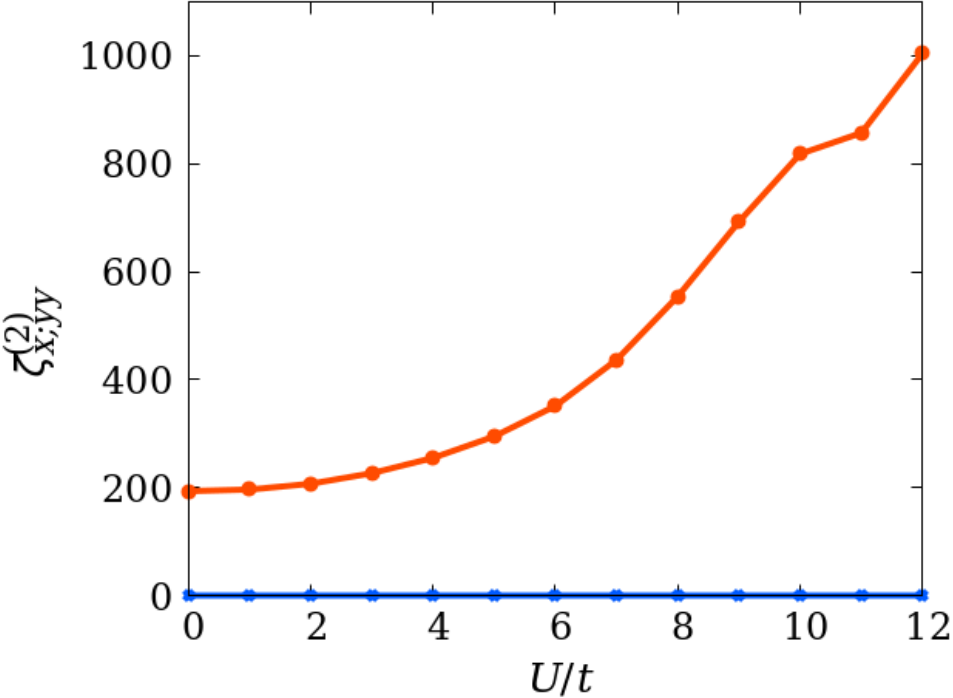}
  \label{fig:CNEE_interaction_U}
  }
  \subfigure[]{
  \includegraphics[height=3.2cm,width=0.468\hsize]{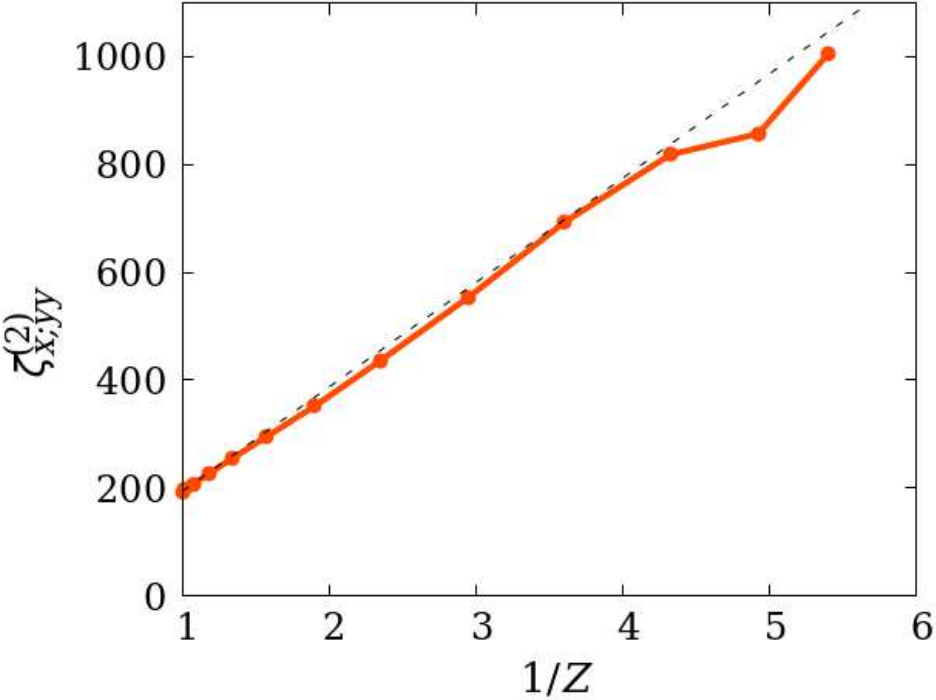}
  \label{fig:CNEE_interaction_Z}
  }
  \setcounter{figure}{2}
  \caption{Magnitude of the calculated current-induced NEE over (a) $U/t$ and (b) $1/Z$. 
  The orange and blue lines in panel (a) indicate the contributions of the Fermi surface terms and the Fermi sea terms, respectively.
  The dashed line in panel (b) corresponds to $\zeta^{(2)}_{x;yy}=Z^{-1}\zeta^{(2),\mathrm{free}}_{x;yy}$.}
  \label{fig:CNEE_interaction}
\end{figure}

\begin{figure*}
  \setcounter{figure}{4}
  \centering
  \subfigure[]{
  \includegraphics[height=4.5cm,width=0.32\hsize]{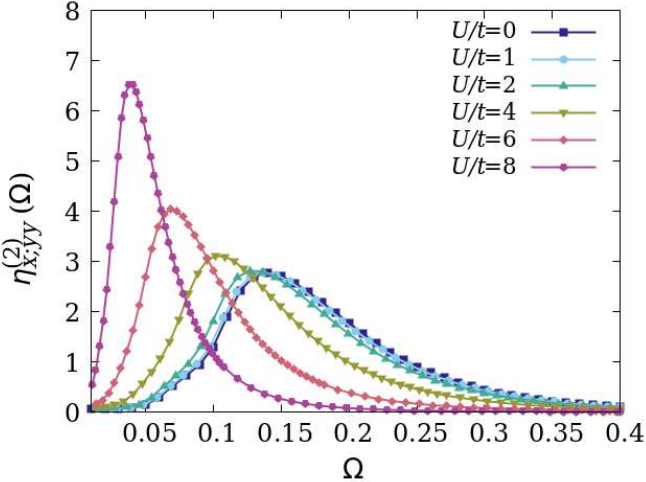}
  \label{fig:LNEE_LPL_interaction_U}
  }
  \subfigure[]{
  \includegraphics[height=4.5cm,width=0.32\hsize]{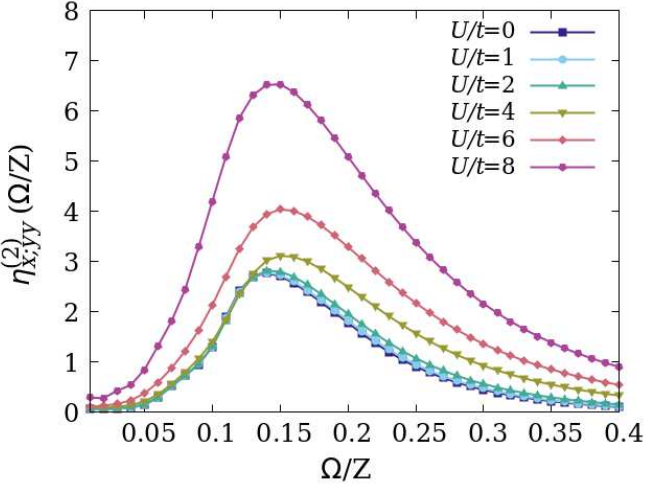}
  \label{fig:LNEE_LPL_interaction_Z}
  }
  \subfigure[]{
  \includegraphics[height=4.5cm,width=0.3\hsize]{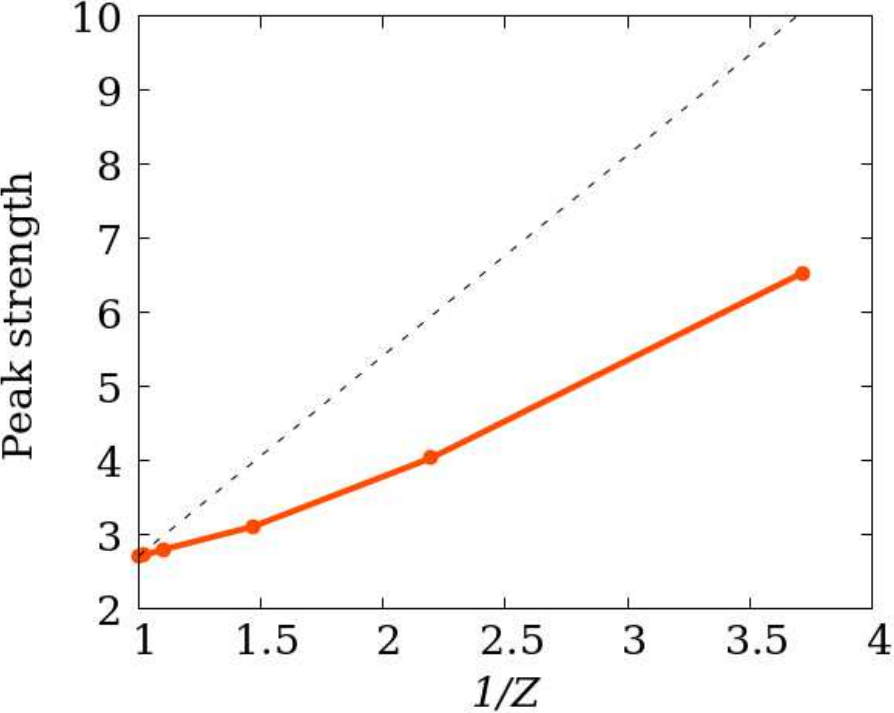}
  \label{fig:peak_strength}
  }
  \setcounter{figure}{3}
  \caption{Calculated LPL-induced NEE over (a) $\Omega$ and (b) $\Omega/Z$ for different interaction strengths. (c) Interaction dependence of the peak height at $\Omega/Z=0.15$.
           The dashed line corresponds to $\eta^{(2)}_{x;yy}(\Omega/Z)=Z^{-1}\eta^{(2),\mathrm{free}}_{x;yy}(\Omega/Z)$ at $\Omega/Z=0.15$ where $\eta^{(2),\mathrm{free}}_{x;yy}=\eta^{(2)}_{x;yy}$ at $U/t=0.0$.}
  \label{fig:LNEE_LPL_interaction}
\end{figure*}


We first consider the symmetry constraints of the NEE. 
The NEE tensor obeys the following symmetry transformation rule:
\begin{align}
  \label{Neumann}
  \zeta^{(2)}_{\alpha';\beta' \gamma'}=\mathrm{det}(\mathcal{R})\mathcal{R}_{\alpha' \alpha}\mathcal{R}_{\beta' \beta}\mathcal{R}_{\gamma' \gamma}\zeta^{(2)}_{\alpha;\beta \gamma}
\end{align}
where $\mathcal{R}$ is a point group operation. 
Then, we divide this response tensor into the symmetric component $\mathcal{S}_{\alpha;\beta \gamma}$ and the antisymmetric component $\mathcal{A}_{\alpha;\beta \gamma}$ regarding the incident electric fields, by defining 
\begin{align}
  \mathcal{S}_{\alpha;\beta \gamma}&=(\zeta^{(2)}_{\alpha;\beta \gamma}+\zeta^{(2)}_{\alpha;\gamma \beta})/2, \\
  \mathcal{A}_{\alpha;\beta \gamma}&=(\zeta^{(2)}_{\alpha;\beta \gamma}-\zeta^{(2)}_{\alpha;\gamma \beta})/2. 
\end{align}
The symmetric part has already been clarified in Ref. \cite{Xiao2023-yu}, which states that
the KMH model belonging to $\mathcal{D}_{3d}$ allows for the following symmetric components: $\mathcal{S}_{x;yy}=\mathcal{S}_{y;xy}=-\mathcal{S}_{x;xx}$.
The antisymmetric part includes $\mathcal{A}_{x;xy}=\mathcal{A}_{y;xy}=0$ and a nonvanishing component, $\mathcal{A}_{z;xy}$.


While the current-induced NEE is a symmetric tensor,
the light-induced NEE can be further classified as the responses under linearly polarized light (LPL) and circularly polarized light (CPL).
According to Ref. \cite{Watanabe2021-kw}, since the electric field in the frequency domain satisfies $\bm{E}(\Omega)=\bm{E}^*(-\Omega)$, 
we can extract the LPL and CPL components from the light-induced NEE
\begin{align}
  \braket{\delta \hat{s}_{\alpha}}=\int \frac{d\Omega}{2\pi} \zeta^{(2)}_{\alpha;\beta \gamma}(\Omega) E^{\beta}(\Omega)E^{\gamma}(-\Omega).
\end{align}
The LPL component can be expressed as
\begin{align}
  \braket{\delta \hat{s}_{\alpha}}_{\mathrm{LPL}}=\int \frac{d\Omega}{2\pi} \eta^{(2)}_{\alpha;\beta \gamma}(\Omega) L^{\beta \gamma}(\Omega),
\end{align}
where we define
\begin{align}
  L^{\beta \gamma}(\Omega)&=\mathrm{Re} \bigl\{ E^{\beta}(\Omega) \bigl[ E^{\gamma}(\Omega)\bigr]^* \bigr\}, \\
  \eta^{(2)}_{\alpha;\beta \gamma}(\Omega)&=\frac{1}{2} \mathrm{Re} \bigl\{ \zeta^{(2)}_{\alpha;\beta \gamma}(\Omega) + \zeta^{(2)}_{\alpha;\gamma \beta}(\Omega) \bigr\}. \label{eta} 
\end{align}
Similarly, the CPL component can be expressed as
\begin{align}
  \braket{\delta \hat{s}_{\alpha}}_{\mathrm{CPL}}=\int \frac{d\Omega}{2\pi} \varepsilon_{\beta \gamma \delta }\xi^{(2)}_{\alpha;\beta \gamma}(\Omega)C^{\delta}(\Omega),
\end{align}
where we define
\begin{align}
  \bm{C}(\Omega)&=\frac{i}{2} \bm{E}(\Omega) \times \bm{E}^*(\Omega), \\
  \xi^{(2)}_{\alpha;\beta \gamma}(\Omega)&=\frac{1}{2} \mathrm{Im} \bigl\{ \zeta^{(2)}_{\alpha;\beta \gamma}(\Omega) - \zeta^{(2)}_{\alpha;\gamma \beta}(\Omega) \bigr\}.  \label{xi} 
\end{align}
Thus, $\eta^{(2)}_{\alpha;\beta \gamma}(\Omega)$ is a symmetric tensor, and $\xi^{(2)}_{\alpha;\beta \gamma}(\Omega)$ is an antisymmetric tensor.


\begin{figure}[b]
  \centering
  \includegraphics[height=4.5cm]{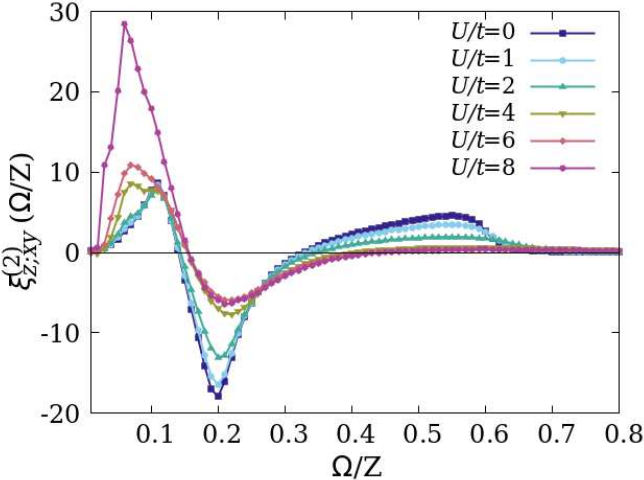}
  \caption{Calculated CPL-induced NEE over $\Omega/Z$ for different interaction strengths.}
  \label{fig:LNEE_CPL_interaction}
\end{figure}

\begin{figure*}
  \setcounter{figure}{6}
  \centering
  \subfigure[]{
  \includegraphics[height=4.5cm,width=0.31\hsize]{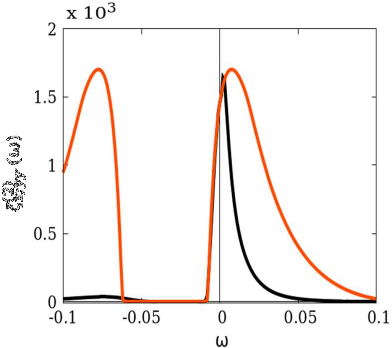}
  \label{fig:CNEE_coh}
  }
  \subfigure[]{
  \centering\includegraphics[height=4.55cm,width=0.30\hsize]{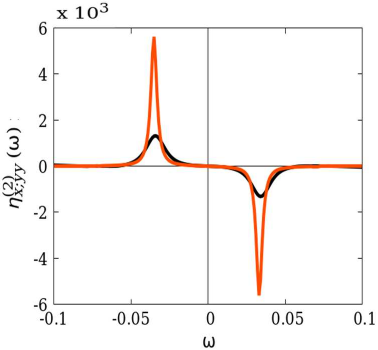}
  \label{fig:LNEE_LPL_coh}
  }
  \subfigure[]{
  \centering\includegraphics[height=4.5cm,width=0.31\hsize]{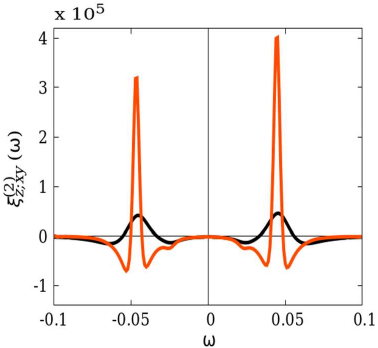}
  \label{fig:LNEE_CPL_coh}
  }
  \setcounter{figure}{5}
  \caption{Frequency-resolved response functions at $U/t=6$ for (a) the current-induced NEE, (b) the LPL-induced NEE ($\Omega/Z=0.15$), and (c) the CPL-induced NEE ($\Omega/Z=0.20$).
  The black and orange lines indicate contributions of the full Green's function and the coherent part of the Green's function, respectively.}
  \label{fig:coh}
\end{figure*}


\subsection{Current-induced NEE} \label{subsec:results-A}

Figure \ref{fig:CNEE_interaction_U} shows the interaction dependence of the current-induced NEE, 
where we separately show the contributions of the Fermi surface terms (orange line) and the Fermi sea terms (blue line) in Eq.~\eqref{CNEE_GR}.
The current-induced NEE requires time-reversal ($\mathcal{T}$) symmetry breaking because the spin is odd but an electric field is even under $\mathcal{T}$.
Dissipation due to impurity scatterings can fulfill this role in nonmagnetic materials. 
However, the dissipation does not appear in the Fermi sea term because the electrons below the Fermi surface are not free to move.
Thus, the contribution of the Fermi sea terms becomes zero, as shown in Fig. \ref{fig:CNEE_interaction_U}, and we ignore these terms in the following.
One can see that the response enhances quadratically in the weakly correlated region and linearly in the strongly correlated region with increasing the interaction.
Furthermore, we replot the results in Fig.~\ref{fig:CNEE_interaction_Z} by replacing the horizontal axis with the inverse of the renormalization factor
\begin{align}
  \label{renormalization factor}
  Z=\biggl(\left. 1- \frac{1}{\hbar} \frac{\partial \mathrm{Re} \Sigma(\omega)}{\partial \omega}\right|_{\omega=0}\biggr)^{-1},
\end{align} 
where $Z^{-1}>1$ for correlated electron systems.
The dashed line represents 
\begin{align}
  \label{Z_inverse}
  \zeta^{(2)}_{x;yy}=Z^{-1}\zeta^{(2),\mathrm{free}}_{x;yy},
\end{align}
where $\zeta^{(2),\mathrm{free}}_{x;yy}=\zeta^{(2)}_{x;yy}$ at $U/t=0.0$.
The shape of the response follows the dashed line, which is consistent with previous results showing that the renormalization effect enhances the nonlinear DC conductivity by $Z^{-1}$
compared to the noninteracting system \cite{Michishita2021-yw,Kofuji2021-ti}.


\subsection{Light-induced NEE} \label{subsec:results-B}

Next, we focus on the light-induced NEE.
First, we show the interaction dependence of the LPL-induced NEE in Fig. \ref{fig:LNEE_LPL_interaction_U}.
There is a single peak in the spectrum where the NEE becomes maximal.
As the interaction increases, the peak position shifts toward the low-frequency region, gradually enhancing the peak strengths.
To analyze the influence of the renormalization, we use the renormalized frequency $\Omega/Z$ to focus on the same interband transitions, which is shown in Fig. \ref{fig:LNEE_LPL_interaction_Z}.
We see that electron correlations also enhance the contribution from these interband transitions. However, the magnitudes are not enhanced by the factor of $Z^{-1}$, as visible in Fig.~\ref{fig:peak_strength}, 
which shows the magnitude of the peak at $\Omega/Z=0.15$ over the renormalization parameter. 
It is clearly visible that the LPL-induced NEE does not follow a linear behavior as $Z^{-1}$. 


More interesting is the interaction dependence of the CPL-induced NEE, shown in Fig. \ref{fig:LNEE_CPL_interaction}.
There are three peaks whose peak positions imply contributions from interband transitions around $K$, $M$, and $\varGamma$ points (of the BZ) in ascending order [see Fig.~\ref{Fig:energy_dispersion}].
Interaction effects are different for these peaks: the peak at the lowest frequency is enhanced by interactions, while the peaks at higher energies are suppressed. 

%% file: sections/section05.tex
\section{Discussion} \label{sec:Discussion}

\begin{figure*}
    \setcounter{figure}{7}
    \centering
    \subfigure[]{
    \includegraphics[height=4.5cm]{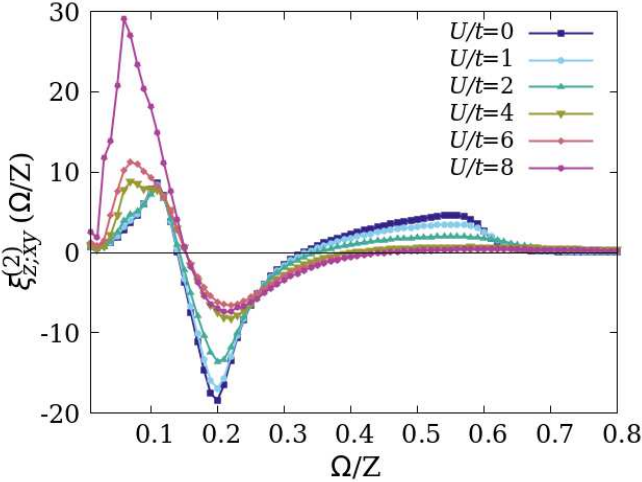}
    \label{fig:LNEE_CPL_d_interaction}
    }
    \hspace{1cm}
    \subfigure[]{
    \centering\includegraphics[height=4.5cm]{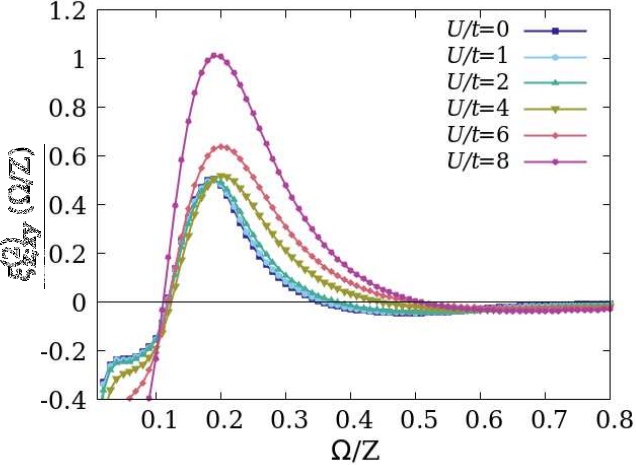}
    \label{fig:LNEE_CPL_o_interaction}
    }
    \subfigure[]{
    \centering\includegraphics[height=4.5cm]{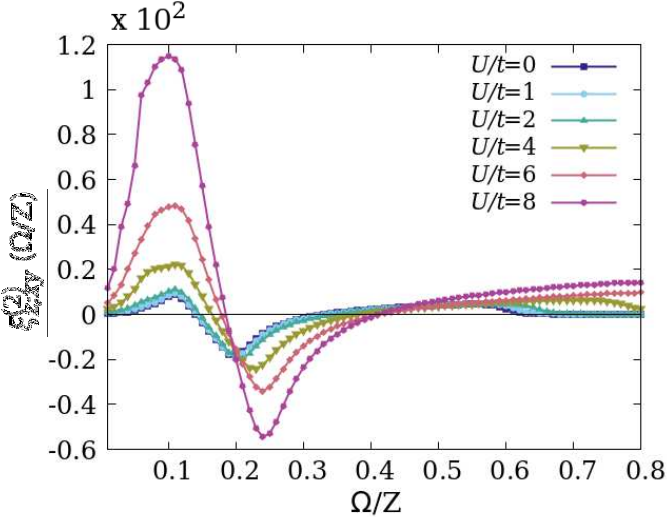}
    \label{fig:LNEE_CPL_d_0ImSigma}
    }
    \hspace{1cm}
    \subfigure[]{
    \centering\includegraphics[height=4.5cm]{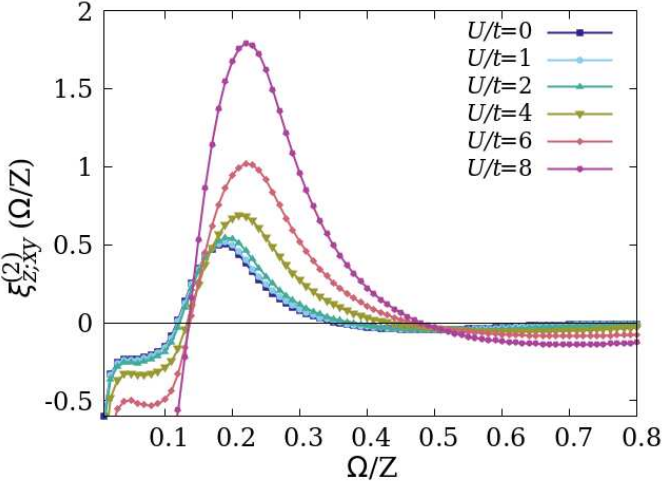}
    \label{fig:LNEE_CPL_o_0ImSigma}
    }
    \setcounter{figure}{6}
    \caption{CPL-induced NEE calculated by including or neglecting the imaginary part of the self-energy. 
    The upper panels show the NEE including the imaginary part for (a) the diagonal part and (b) the off-diagonal part.
    The lower panels show the NEE neglecting the imaginary part for (c) the diagonal part and (d) the off-diagonal part.}
    \label{fig:LNEE_CPL_d/o}
\end{figure*}


Based on the results of the previous section,
we first analytically show that the renormalization effect enhances the current-induced NEE by $Z^{-1}$, similar to the nonlinear DC conductivity, in Sec. \ref{subsec:Discussion-A}.
Furthermore, we discuss the weak degree of enhancement for the light-induced NEE.
Then, we explain the complex frequency dependence exhibited by the CPL-induced NEE in Sec. \ref{subsec:Discussion-B}.
Finally, we note that the light-induced NEE is closely related to the photomagnetic and optomagnetic effects in Sec. \ref{subsec:Discussion-C}.


\subsection{Origin of the increased response and differences in the degree of enhancement} \label{subsec:Discussion-A}

The current-induced response is mainly generated at the Fermi energy ($\omega=0$).
The self-energy can be expanded in the vicinity of the Fermi energy as
\begin{align}
    \label{frequency_expansion}
    \Sigma^R(\omega) \sim \left. \mathrm{Re}\Sigma^R(0)+\hbar \omega \frac{\partial \mathrm{Re} \Sigma^R(\omega)}{\hbar \partial \omega} \right|_{\omega=0}+i\mathrm{Im}\Sigma^R(0). 
\end{align}
This expansion allows us to divide the Green's function into a coherent part $G^R_{\mathrm{coh}}$ and an incoherent part $G^R_{\mathrm{inc}}$ as 
\begin{align}
    G^R(\omega)&=G^R_{\mathrm{coh}}(\omega)+G^R_{\mathrm{inc}}(\omega), \\
    G^R_{\mathrm{coh}}(\omega)&=\frac{Z}{\hbar \omega-Z\varepsilon'(\bm{k})+i\Gamma},
\end{align}
where $\varepsilon'(\bm{k})=\varepsilon(\bm{k})-\mu+\mathrm{Re}\Sigma(0)$ and $\Gamma=-Z\mathrm{Im}\Sigma(0)$.
The coherent part expresses a quasiparticle with energy $\varepsilon'(\bm{k})$ and lifetime $\hbar/\Gamma$.
We note that $\Gamma$ is very small in our calculations.
Supposing that the Green's function can be expressed only by its coherent part and that $\Gamma$ is sufficiently small, 
we can describe the Green's function as 
\begin{align}
    G^R(\omega)=G^{R,\mathrm{free}}(Z^{-1}\omega),
\end{align}
where $G^{R,\mathrm{free}}$ is the Green's function of the noninteracting system.
Following the same procedure as Ref. \cite{Michishita2021-yw}, we can find
\begin{align}
    \zeta^{(2)}_{\alpha;\beta \gamma} \simeq Z^{-1} \zeta^{(2),\mathrm{free}}_{\alpha;\beta \gamma}, 
\end{align}
which states that the renormalization effect enhances the current-induced NEE by $Z^{-1}$.
Obviously, this is an effect of the real part of the self-energy.
On the other hand, the expansion of Eq. \eqref{frequency_expansion} is not valid for the light-induced NEE due to the frequency dependence of the self-energy.
Nevertheless, we cannot explain the increased response for the light-induced NEE without the renormalization effect.
Therefore, to evaluate the contribution of the coherent part, we calculate the frequency-resolved response function $\zeta^{(2)}_{\alpha;\beta \gamma}(\omega)$ which is given as
\begin{eqnarray}
    \zeta^{(2)}_{\alpha;\beta \gamma}=\int \frac{d \omega}{2 \pi} F(\omega) \zeta^{(2)}_{\alpha;\beta \gamma}(\omega),
\end{eqnarray}
where $F(\omega)=-\partial f(\omega)/\partial \omega$ for the current-induced NEE and $F(\omega)=f(\omega)$ for the light-induced NEE.


Figure \ref{fig:CNEE_coh} shows the frequency-resolved response function for the current-induced NEE, comparing the response induced by the full Green's function (black line) with the response induced by the coherent part of the Green's function (orange line).
The contribution of the coherent part near the Fermi energy is identical to the total contribution.
On the other hand, for the light-induced NEE, the comparisons in Figs.~\ref{fig:LNEE_LPL_coh} and \ref{fig:LNEE_CPL_coh} between the responses induced by the full Green's function and its coherent part demonstrates that the coherent part 
overestimates the response.
Thus, the light-induced NEE is not as strongly enhanced as the current-induced NEE due to the frequency dependence of the self-energy.


\subsection{Complex frequency dependence exhibited by the CPL-induced NEE} \label{subsec:Discussion-B}

We analyze the CPL-induced NEE by using the Bloch basis and Eq.~\eqref{spin_in_Bloch}.
We first divide Eq. \eqref{Second-order response function} into the diagonal part ($\propto s^{nn}_\alpha(\bm{k})$) and the off-diagonal part ($\propto s^{nm}_\alpha(\bm{k})$). 
Figures \ref{fig:LNEE_CPL_d_interaction} and \ref{fig:LNEE_CPL_o_interaction} show the contribution of the diagonal and off-diagonal parts to the CPL-induced NEE, respectively.
Although the main contribution comes from the diagonal part, the response from the off-diagonal part is also finite.
Most importantly, electron correlations enhance the off-diagonal response, while the response from the diagonal part is only enhanced for small frequencies.
We note that the diagonal response is an intraband effect that depends on an extrinsic scattering time, while the off-diagonal part is an interband effect that is intrinsic and only depends on the light frequency, which can be confirmed in Eq.~(54) in Ref.~\cite{Fregoso2022-fd}. 
Furthermore, the LPL-induced NEE is also an interband effect, as seen in Eq.~(53) in Ref.~\cite{Fregoso2022-fd}.
Therefore, a reduction of the lifetime due to electron correlations can suppress the dissipative response, such as the diagonal part of the CPL response.
To investigate this hypothesis, we recalculate the CPL-induced NEE by effectively excluding the imaginary part of the self-energy reflecting dissipation.


Figure \ref{fig:LNEE_CPL_d_0ImSigma} shows the result for the diagonal part. 
The second peak is indeed enhanced when neglecting the imaginary part of the self-energy, which indirectly shows that the imaginary part of the self-energy suppresses the response.
Besides the suppression by the imaginary part, the diagonal part can be enhanced by the real part of the self-energy, which creates a complex frequency dependence.
For example, the first peak in the low-frequency region is enhanced because the renormalization effect of the real part exceeds the imaginary part.
On the other hand, for the peaks at larger frequencies, the suppression by the imaginary part is stronger [see Fig. \ref{fig:LNEE_CPL_d_interaction}].


Furthermore, the frequency dependence of the real part of the self-energy can explain why the position of the low-frequency peak shifts to smaller frequencies.
In particular, the renormalization effect is weaker for larger frequencies.
Together with the imaginary part of the self-energy, this leads to a shift in the peak position for large interaction strengths [compare Figs. \ref{fig:LNEE_CPL_d_interaction} and \ref{fig:LNEE_CPL_d_0ImSigma}].
On the other hand, the off-diagonal response does not include a strong effect of the imaginary part of the self-energy.
Comparing between Fig.~\ref{fig:LNEE_CPL_o_interaction} and \ref{fig:LNEE_CPL_o_0ImSigma}, we see that the real part of the self-energy is sufficient to explain the NEE.


\subsection{Relationship between the light-induced NEE and photomagnetic or optomagnetic effects} \label{subsec:Discussion-C}

The photomagnetic and optomagnetic effects are non-thermal phenomena where light changes the magnetic properties of a material 
but does not involve the heating of electrons by the laser pulse \cite{Kirilyuk2010-ys}.
The photomagnetic effects depend on the absorption of photons and are related to the optical spin injection \cite{Oestreich2002-ax,Utic2004-za}.
The optomagnetic effects do not require the absorption of photons but instead are referred to as the IFE \cite{Kimel2005-ir} and the inverse Cotton-Mouton effect (ICME) \cite{Kalashnikova2007-yk}.


The optical spin injection, which is sometimes called the optical orientation, generates a finite spin expectation value of the electrons by transferring the angular momentum of the photons during an optical excitation with CPL.
The optical spin injection, for example, has been used for exciting the persistent spin helix in semiconductor quantum wells \cite{Koralek2009-hw,Walser2012-jr}.
Using the band representation \cite{Xu2021-hx,Fregoso2022-fd,Nastos2007-jn}, we can show that the diagonal part of the CPL-induced NEE corresponds to the optical spin injection.
Therefore, we can conclude that electron correlations can either enhance or suppress the optical spin injection depending on the light frequency, as shown in Fig. \ref{fig:LNEE_CPL_d_interaction}.


On the other hand, the IFE and ICME generate an effective magnetic field under CPL and LPL, respectively, which can be described following Ref. \cite{Xu2021-hx} as
\begin{align}
    H_{\mathrm{IFE}}^\gamma &\propto a_{\alpha \beta \gamma} \bigl\{E^\alpha(\Omega)[E^\beta(\Omega)]^*-E^\beta(\Omega)[E^\alpha(\Omega)]^* \bigr\},\label{IFE} 
\end{align}
\begin{align}    
    H_{\mathrm{ICME}}^\gamma &\propto b_{\alpha \beta \gamma \delta} M^\delta  \bigl\{E^\alpha(\Omega)[E^\beta(\Omega)]^*+E^\beta(\Omega)[E^\alpha(\Omega)]^* \bigr\}, \label{ICME}
\end{align}
where $\bm{H}_{\mathrm{IFE}/\mathrm{ICME}}$ is the effective magnetic field for the IFE/ICME, $\bm{M}$ is a DC magnetization, and $a_{\alpha \beta \gamma}$ and $b_{\alpha \beta \gamma \delta}$ are phenomenological parameters.
The IFE and ICME look at the same effects as the light-induced NEE in the sense that the spin can be controlled by an effective magnetic field.
These responses require $\mathcal{T}$-breaking elements similar to the current-induced NEE.


For the LPL-induced NEE, energy dissipation associated with interband transitions plays the role of the $\mathcal{T}$-breaking element,
which allows us to observe this effect even in nonmagnetic materials ($M=0$).
On the other hand, the ICME is finite only in magnets ($M\neq0$) from Eq. \eqref{ICME}.
This is because the ICME is an equilibrium phenomenon, while the LPL-induced NEE is a nonequilibrium response.
Therefore, from this study, we cannot conclude whether electron correlations can enhance or suppress the ICME. The ICME must be analyzed in magnets to study the relationship between the ICME and electron correlations.


Unlike LPL, CPL creates a $\mathcal{T}$-symmetry breaking external field because the helicity of the light changes under $\mathcal{T}$.
In other words, both the IFE and the CPL-induced NEE are active in nonmagnetic materials. 
Notably, the IFE corresponds to the off-diagonal part of the CPL-induced NEE 
because the off-diagonal part is nonzero even in the nonresonant-frequency region ($\Omega/Z \ll 0.10$), as shown in Fig. \ref{fig:LNEE_CPL_o_interaction},
which shows that it does not depend on the absorption of photons.
Therefore, we can conclude from Fig. \ref{fig:LNEE_CPL_o_interaction} that electron correlations can enhance the IFE regardless of the light frequency.


\begin{table}[b]
    \caption{Response characteristics on the NEE to incident electric fields:
    The real and imaginary parts of the self-energy can enhance ($\nearrow$) or suppress ($\searrow$) the NEE, respectively.
    ``$\checkmark/\times$'' indicates whether the effect exists or not. We use 
    ``{$\checkmark$}\llap{$\diagdown$}'' if the effect is possible but not as large as ``$\checkmark$''.}
    \vspace{1mm}
    \label{tab:summary}
    \centering
    \tabcolsep = 1.4mm
    \renewcommand\arraystretch{1.3}
    \begin{tabular}{cc|ccc}
      \hline \hline
      \multirow{2}{*}{Self-energy} & Effect on  & \multirow{2}{*}{Current-induced} & \multicolumn{2}{c}{\multirow{2}{*}{Light-induced}}  \\
      & the response & & & \\
      \hline 
      & & & LPL & CPL \\
      Real & $\nearrow$ & $\checkmark$ & {$\checkmark$}\llap{$\diagdown$} & {$\checkmark$}\llap{$\diagdown$}  \\
      \hline
      Imaginary & $\searrow$ & $\times$ & $\times$ & $\checkmark$  \\
      \hline \hline
    \end{tabular}
    \vspace{5mm}
\end{table}

%% file: sections/section06.tex
\section{summary and future outlook} \label{sec:Summary and future outlook}

We have derived a formalism for the nonlinear Edelstein effect (NEE) with a full quantum mechanical approach and
analyzed the impact of electron correlations on the current-induced NEE and the light-induced NEE by performing numerical calculations on a Hubbard model.
The light-induced NEE is further classified as the responses under linearly polarized light (LPL) and circularly polarized light (CPL).
We have found that electron correlations can either enhance or suppress nonlinear responses.
The enhancement of the response is due to the renormalization effect, which originates in the real part of the self-energy.
On the other hand, the suppression can only occur in dissipative responses and depends on the imaginary part of the self-energy.
In other words, the real part of the self-energy enhances the NEE, and the imaginary part suppresses it.
Specifically, the current-induced NEE and the LPL-induced NEE include only the real-part effect, while the CPL-induced NEE includes both real-part and imaginary-part effects because it has a dissipative intraband contribution.
However, we found that the light-induced NEE is not as strongly enhanced as the current-induced NEE due to the frequency dependence of the self-energy.
Finally, table~\ref{tab:summary} summarizes the effects of the self-energy on different types of the NEE.


In this work, we have systematically analyzed the light-induced NEE which includes photomagnetic or optomagnetic effects as specific cases.
Notably, the obtained formulas are based on Green's functions, which allows us to investigate these effects in strongly correlated electron systems (SCES).
We have focused on the optical spin injection for the photomagnetic effects, and the inverse Faraday effect (IFE) and the inverse Cotton-Mouton effect (ICME) for the optomagnetic effects.
The optical spin injection is the diagonal response of the CPL-induced NEE, and it can be either enhanced or suppressed depending on the light frequency due to the competition between the real-part and imaginary-part effects.
On the other hand, the IFE is the off-diagonal response, which is enhanced because of the absence of the imaginary-part effect.
We stress that the above analyses focused on insulators. The light-induced NEE in metals has the other intraband effect originating from the Fermi surface contribution \cite{Fregoso2022-fd}.
Therefore, the IFE in metals would show a more complex behavior.
In addition, we have not discussed the impact of electron correlations on the ICME in this study
because it is an equilibrium phenomenon that is inactive in nonmagnetic materials, while the LPL-induced NEE is a nonequilibrium phenomenon that is active even in them.
We will leave these open questions as future work.


Recently, T. Kodama \textit{et al}. have experimentally detected the current-induced nonlinear magnetoelectric effect on Pt-Py bilayers \cite{Kodama2023-ev}.
Their study suggests that one can experimentally capture the current-induced NEE in SCES.
In addition, the equations formulated in this study enable quantitative evaluation by combining first-principles calculations with dynamical mean-field theory, 
which will give material platforms for ``strongly correlated spintronics" mediated by nonlinear responses.

%% file: appendices/appendixA.tex
\section{DERIVATION OF THE RESPONSE FUNCTION BASED ON THE PATH INTEGRAL MATSUBARA FORMALISM} \label{app:A}

The derivation of Eq.~\eqref{Second-order response function} in the main text consists of four main steps. 
We also derive the first-order response function for reference and omit  $\int d\bm{k}/(2\pi)^d$.


First, we express the response functions in terms of Matsubara Green's functions. The response functions can be expressed by functional derivatives as
\begin{align}
\chi^{(1)}_{\alpha;\beta}(\tau;\tau_1)=&\left. \frac{1}{Z[0]} \frac{\delta}{\delta A^{\beta}(\tau_1)} \frac{\delta}{\delta B^{\alpha}(\tau)}\right|_{B=A=0}Z[A,B]  \\
=& \sum_{\lambda,\eta,\sigma,\rho} \braket{ \bar{\psi}_\lambda(\tau) s^{\lambda \eta}_{\alpha} \psi_{\eta}(\tau) \bar{\psi}_{\sigma}(\tau_1)J^{\sigma \rho}_{\beta} \psi_{\rho}(\tau_1)}, \\
\chi^{(2)}_{\alpha;\beta \gamma }(\tau;\tau_1,\tau_2)=&\left. \frac{1}{Z[0]} \frac{\delta}{\delta A^{\gamma}(\tau_2)}\frac{\delta}{\delta A^{\beta}(\tau_1)} \frac{\delta}{\delta B^{\alpha}(\tau)}\right|_{B=A=0}Z[A,B] \\
=& -\delta(\tau_1-\tau_2) \sum_{\lambda,\eta,\sigma,\rho} \braket{ \bar{\psi}_\lambda(\tau) s^{\lambda \eta}_{\alpha} \psi_{\eta}(\tau) \bar{\psi}_{\sigma}(\tau_1)J^{\sigma \rho}_{\beta \gamma} \psi_{\rho}(\tau_1)} \notag  \\
&+\sum_{\lambda,\eta,\sigma,\rho,\mu,\nu} \braket{ \bar{\psi}_\lambda(\tau) s^{\lambda \eta}_{\alpha} \psi_{\eta}(\tau) \bar{\psi}_{\sigma}(\tau_1)J^{\sigma \rho}_{\beta} \psi_{\rho}(\tau_1) \bar{\psi}_{\mu}(\tau_2)J^{\mu \nu}_{\gamma} \psi_{\nu}(\tau_2) } ,
\end{align}
where $\braket{X}=Z[0]^{-1}\int \mathcal{D}\bar{\psi} \mathcal{D}\psi X e^{-S[0]}$ is the functional integral over the action without external fields.
Using Wick’s theorem and neglecting vertex corrections, we can write these many-particle Green's functions as products of single-particle Green's functions and obtain
\begin{align}
\chi^{(1)}_{\alpha;\beta}(\tau;\tau_1)&=-\sum_{\lambda,\eta,\sigma,\rho} s^{\lambda \eta}_{\alpha}\braket{-\psi_{\eta}(\tau)\bar{\psi}_{\sigma}(\tau_1)}J^{\sigma \rho}_{\beta} \braket{-\psi_{\rho}(\tau_1)\bar{\psi}_{\lambda}(\tau)}, \label{A5} \\
\chi^{(2)}_{\alpha;\beta \gamma }(\tau;\tau_1,\tau_2)&= \delta(\tau_1-\tau_2 ) \sum_{\lambda,\eta,\sigma,\rho}  s^{\lambda \eta}_{\alpha} \braket{ -\bar{\psi}_\eta(\tau) 
\psi_{\sigma}(\tau_1)} J^{\sigma \rho}_{\beta \gamma}  \braket{-\bar{\psi}_{\rho}(\tau_1) \psi_{\lambda}(\tau)} \notag  \\
+ \sum_{\lambda,\eta,\sigma,\rho,\mu,\nu} \Bigl(  &s^{\lambda \eta}_{\alpha}  \braket{ -\psi_\eta(\tau)  \bar{\psi}_{\sigma}(\tau_1)} J^{\sigma \rho}_{\beta} \braket{-\psi_{\rho}(\tau_1) \bar{\psi}_{\mu}(\tau_2)} J^{\mu \nu}_{\gamma} \braket{-\psi_{\nu}(\tau_2) \bar{\psi}_{\lambda}(\tau) }
+ \bigl[(\beta,\tau_1)\leftrightarrow (\gamma,\tau_2) \bigr] \Bigr).  \label{A6}
\end{align}
Note that the other terms vanish because they are proportional to the expectation value of a current operator without an applied electric field.


Second, we take the Fourier transformation and derive the response functions in terms of the Matsubara frequencies as follows:
\begin{align}
\chi^{(1)}_{\alpha;\beta}(i\omega_n;i\omega_n)
=&-\frac{1}{\beta} \sum_{\omega_l} \mathrm{Tr} \bigl[ s_{\alpha}\mathscr{G} (i\omega_l+i\omega_n) J_{\beta} \mathscr{G} (i\omega_l) \bigr], \\
\chi^{(2)}_{\alpha;\beta \gamma} (i\omega_n+i\omega_m;i\omega_n,i\omega_m)
=&\frac{1}{\beta} \sum_{\omega_l} \mathrm{Tr} \biggl[  \frac{1}{2} s_{\alpha} \mathscr{G}(i\omega_l+i\omega_n+i\omega_m) J_{\beta\gamma} \mathscr{G}(i\omega_l) \notag \\
&+ s_{\alpha} \mathscr{G}(i\omega_l+i\omega_n+i\omega_m) J_{\beta} \mathscr{G}(i\omega_l+i\omega_m) J_{\gamma} \mathscr{G}(i\omega_l)\biggr]+\bigl[(\beta,i\omega_n)\leftrightarrow (\gamma,i\omega_m) \bigr], 
\end{align}
where $\omega_l=(2l+1)\pi/\beta$ is a Fermionic Matsubara frequency, and $\omega_n=2n\pi/\beta$ and $\omega_m=2m\pi/\beta$ are Bosonic Matsubara frequencies which originate from photons.
Here, we define $\mathscr{G}(\tau-\tau')=\braket{-\psi(\tau)\bar{\psi}(\tau')}$, and
use $\mathscr{G}(\tau)=\beta^{-1}\sum_l \mathscr{G}(i\omega_l)e^{-i\omega_l \tau}$ and $\int^{\beta}_{0} e^{i(\omega_n-\omega_m)\tau}=\beta \delta_{nm}$.


Third, we perform the sum over the Matsubara frequencies by using the identity $\beta^{-1}\sum_l X(i \omega_l)=-\oint_C\frac{dz}{2\pi i} f(z)X(z)$ where $\oint_C$ represents paths only around the poles of the Fermi distribution function $f(z)=(e^{\beta z}+1)^{-1}$ avoiding poles of $X(z)$, as shown in Fig. \ref{fig:Matsubara_sum1}. 
Since each path can be transformed within a regular region [Fig. \ref{fig:Matsubara_sum2}], the response functions can be calculated by
\begin{align}
&\chi^{(1)}_{\alpha;\beta}(i\omega_n;i\omega_n)
=\int^{\infty}_{-\infty} \frac{d\varepsilon}{2\pi i}f(\varepsilon) \notag \\
&\hspace{2.5cm}\times \mathrm{Tr} \bigl[ s_{\alpha} \mathscr{G}(\varepsilon+i\omega_n)J_{\beta}(\mathscr{G}(\varepsilon+i\eta)-\mathscr{G}(\varepsilon-i\eta))+s_{\alpha}(\mathscr{G}(\varepsilon+i\eta)-\mathscr{G}(\varepsilon-i\eta))J_{\beta} \mathscr{G}(\varepsilon-i\omega_n)\bigr], \\
&\chi^{(2)}_{\alpha;\beta \gamma}(i\omega_n+i\omega_m;i\omega_n,i\omega_m) 
=-\int^{\infty}_{-\infty} \frac{d\varepsilon}{2 \pi i} f(\varepsilon) \notag \\
&\times\mathrm{Tr} \biggl[\frac{1}{2}  \Bigl( s_{\alpha} \mathscr{G}(\varepsilon+i\omega_n+i\omega_m)J_{\beta\gamma}(\mathscr{G}(\varepsilon+i\eta)-\mathscr{G}(\varepsilon-i\eta))+s_{\alpha}(\mathscr{G}(\varepsilon+i\eta)-\mathscr{G}(\varepsilon-i\eta))J_{\beta \gamma}\mathscr{G}(\varepsilon-i\omega_n-i\omega_m) \Bigr) \notag \\
 &+s_{\alpha} \mathscr{G}(\varepsilon+i\omega_n+i\omega_m)J_{\beta} \mathscr{G}(\varepsilon+i\omega_m)J_{\gamma}(\mathscr{G}(\varepsilon+i\eta)-\mathscr{G}(\varepsilon-i\eta))+s_{\alpha} \mathscr{G}(\varepsilon+i\omega_n)J_{\beta}(\mathscr{G}(\varepsilon+i\eta)-\mathscr{G}(\varepsilon-i\eta))J_{\gamma} \mathscr{G}(\varepsilon-i\omega_m)\notag \\
 &+s_{\alpha}(\mathscr{G}(\varepsilon+i\eta)-\mathscr{G}(\varepsilon-i\eta))J_{\beta}\mathscr{G}(\varepsilon-i\omega_n)J_{\gamma}\mathscr{G}(\varepsilon-i\omega_n-i\omega_m) \biggr]+\bigl[(\beta,i\omega_n)\leftrightarrow (\gamma,i\omega_m) \bigr]. 
\end{align}
Here, we use $f(\varepsilon-i\omega_n)=f(\varepsilon-i\omega_n-i\omega_m)=f(\varepsilon)$ and $i\omega_n,i\omega_m \gg i \eta$.\par

\begin{figure}[t]
    \centering
    \setcounter{figure}{8}
    \subfigure[]{
    \includegraphics[height=4.5cm,width=0.3\hsize]{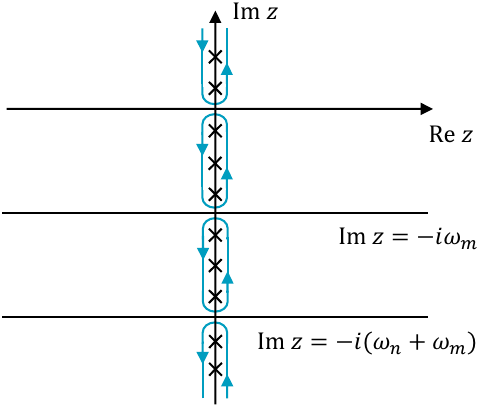}
    \label{fig:Matsubara_sum1}
    }
    \hspace{10mm}
    \subfigure[]{
    \includegraphics[height=4.6cm,width=0.25\hsize]{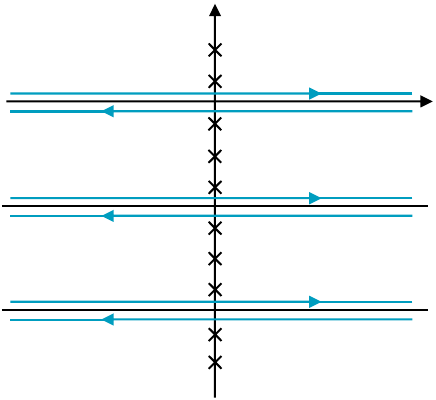}
    \label{fig:Matsubara_sum2}
    } 
    \setcounter{figure}{7}
    \caption{(a) Locations of the poles ($\times$) of $f(z)$ and the integral paths (blue curves) around them. 
    Each integral path does not span different regions of the analyticity of $X(z)$. 
    (b) Integral paths used to sum over the Matsubara frequencies. 
    Each integral path is along the real axis at a distance of $\pm i\eta$ from the analytical boundary of $X(z)$.
    Note that these paths correspond to the case of a second-order response function as $X(z)$.}
    \label{fig:Matsubara_sum}
\end{figure}


Fourth, we perform the analytic continuation by $i\omega_n \rightarrow \hbar\omega_1+i\eta$, $i\omega_m \rightarrow \hbar\omega_2+i\eta$.
By using that analytic functions in the upper/lower plane become the retarded/advanced Green's functions, we can describe the response functions as
\begin{align}
&\chi^{(1)}_{\alpha;\beta}(\omega_1;\omega_1) \notag \\
&=\int^{\infty}_{-\infty} \frac{d\varepsilon}{2\pi i} f(\varepsilon) \mathrm{Tr} \bigl[ s_{\alpha}G^R(\varepsilon+\hbar \omega_1)J_{\beta}(G^R(\varepsilon)-G^A(\varepsilon)) +  s_{\alpha}(G^R(\varepsilon)-G^A(\varepsilon))J_{\beta} G^A(\varepsilon-\hbar \omega_1)\bigr] , \label{A13} \\
&\chi^{(2)}_{\alpha;\beta,\gamma}(\omega_\Sigma;\omega_1,\omega_2) \notag \\
&=-\int^{\infty}_{-\infty} \frac{d\varepsilon}{2\pi i} f(\varepsilon) \biggl\{ \frac{1}{2}\mathrm{Tr} \bigl[ s_{\alpha}G^R(\varepsilon+\hbar\omega_\Sigma)J_{\beta \gamma}(G^R(\varepsilon)-G^A(\varepsilon))+s_{\alpha}(G^R(\varepsilon)-G^A(\varepsilon))J_{\beta \gamma}G^A(\varepsilon-\hbar \omega_\Sigma) \bigr] \notag \\
&\hspace{0.4cm}+\mathrm{Tr} \bigl[ s_{\alpha}G^R(\varepsilon+\hbar\omega_\Sigma)J_{\beta} G^R(\varepsilon+\hbar \omega_2)J_{\gamma} (G^R(\varepsilon)-G^A(\varepsilon))+s_{\alpha} G^R(\varepsilon+\hbar \omega_1) J_{\beta} (G^R(\varepsilon)-G^A(\varepsilon)) J_{\gamma}G^A(\varepsilon-\hbar\omega_2) \notag \\
&\hspace{0.4cm}+s_{\alpha}(G^R(\varepsilon)-G^A(\varepsilon))J_{\beta}G^A(\varepsilon-\hbar \omega_1)J_{\gamma}G^A(\varepsilon-\hbar \omega_\Sigma) \bigr] \biggr\} +\bigl[(\beta,\omega_1)\leftrightarrow (\gamma,\omega_2) \bigr],  
\end{align}
where $\omega_\Sigma=\omega_1+\omega_2$. Using $\zeta^{(2)}_{\alpha;\beta,\gamma}(\omega_\Sigma;\omega_1,\omega_2)=\chi^{(2)}_{\alpha;\beta,\gamma}(\omega_\Sigma;\omega_1,\omega_2)/i(\omega_1+i\delta)i( \omega_2+i\delta) $ and changing the argument from $\varepsilon$ to $\omega$ by $\varepsilon=\hbar\omega$, we can derive Eq. \eqref{Second-order response function} in the main text.

%% file: appendices/appendixB.tex
\section{DC LIMIT} \label{app:B}

We first expand the single-particle Green's function and the Fermi distribution function for small frequencies as follows:
\begin{align}
G^{R/A}(\omega+\omega_1)&\simeq G^{R/A}(\omega)+\frac{\partial G^{R/A}(\omega)}{ \partial \omega} \omega_1, \\
G^{R/A}(\omega+\omega_1+\omega_2)&\simeq G^{R/A}(\omega)+\frac{\partial G^{R/A}(\omega)}{\partial \omega} (\omega_1+\omega_2)+\frac{\partial^2 G^{R/A}(\omega)}{\partial \omega^2}\omega_1\omega_2 ,\\
f(\omega+\omega_1)&\simeq f(\omega)+\frac{\partial f(\omega)}{\partial \omega} \omega_1.
\end{align}
Using these equations, we can rewrite Eq \eqref{Second-order response function} as
\begin{align}
\zeta^{(2)}_{\alpha;\beta\gamma}(\omega_1+\omega_2;\omega_1,\omega_2)=\frac{\hbar}{\omega_1 \omega_2 }\int^{\infty}_{-\infty} \frac{d \omega}{2\pi i} \Bigl\{ A_0(\omega)+A_1(\omega)\omega_1+A'_1(\omega)\omega_2+A_2(\omega)\omega_1 \omega_2 +O(\omega_i^3)\Bigr\},   
\end{align}
where each component is given by
\begin{align} 
&A_0(\omega)=\int \frac{d\bm{k}}{(2\pi)^d} f(\omega) \mathrm{Tr} \biggl[ \frac{1}{2} \Bigl( s_{\alpha}G^R(\omega)J_{\beta \gamma}G^R(\omega)-s_{\alpha}G^A(\omega)J_{\beta \gamma}G^A(\omega) \Bigr) \notag \\
& \hspace{1.3cm} +s_{\alpha}G^R(\omega)J_{\beta} G^R(\omega)J_{\gamma} G^R(\omega)- s_{\alpha}G^A(\omega)J_{\beta}G^A(\omega)J_{\gamma}G^A(\omega) \biggr] + (\beta \leftrightarrow \gamma), \label{B5}\\
&A_1(\omega)= \int \frac{d\bm{k}}{(2\pi)^d} \biggl(\frac{\partial f(\omega)}{\partial \omega}\biggr) \mathrm{Tr} \bigl[s_{\alpha}G^R(\omega)J_{\beta \gamma} G^A(\omega)+s_{\alpha}G^R(\omega)J_{\beta}G^A(\omega)J_{\gamma}G^A(\omega) + s_{\alpha}G^R(\omega)J_{\gamma}G^R(\omega)J_{\beta}G^A(\omega)\bigr] 
 \notag \\
& \hspace{1.3cm} +\int \frac{d \bm{k}}{(2\pi)^d} f(\omega)\biggl\{ \mathrm{Tr} \biggl[s_{\alpha} \frac{\partial G^R(\omega)}{\partial \omega}J_{\beta \gamma}G^R(\omega)  + s_{\alpha} \frac{\partial G^R(\omega)}{\partial \omega}J_{\beta} G^R(\omega)J_{\gamma} G^R(\omega) \notag \\
& \hspace{4.2cm}+s_{\alpha}\frac{\partial G^R(\omega)}{\partial \omega}J_{\gamma} G^R(\omega)J_{\beta} G^R(\omega)+s_{\alpha}G^R(\omega)J_{\gamma}\frac{\partial G^R(\omega)}{\partial \omega}J_{\beta}G^R(\omega) \biggr]+c.c.\biggr\}, \label{B6}\\
&A'_1(\omega)= A_1(\omega;\beta \leftrightarrow \gamma), \\
&A_2(\omega)=\int \frac{d\bm{k}}{(2\pi)^d}\biggl(\frac{\partial f(\omega)}{\partial \omega}\biggr) \biggl\{ \mathrm{Tr} \biggl[ s_{\alpha}\frac{\partial G^R(\omega)}{\partial \omega} J_{\beta}G^R(\omega)J_{\gamma}G^A(\omega)+\frac{1}{2} s_{\alpha} \frac{\partial G^R(\omega)}{\partial \omega}J_{\beta \gamma} G^A(\omega)\biggr]  -c.c. \biggr\} \notag \\
&\hspace{1.3cm} + \int \frac{d\bm{k}}{(2\pi)^d} f(\omega) \biggr\{ \mathrm{Tr} \biggl[ s_{\alpha} \frac{\partial}{\partial \omega} \biggl( \frac{\partial G^R(\omega)}{\partial \omega} J_{\beta}G^R(\omega)\biggr)J_{\gamma} G^R(\omega)+ \frac{1}{2} s_{\alpha} \frac{\partial^2 G^R(\omega)}{\partial \omega^2}J_{\beta \gamma}G^R(\omega) \biggr] -c.c. \biggl\}  +(\beta \leftrightarrow \gamma).  
\end{align}
Here, we use $(G^R)^\dagger=G^A$ when calculating $c.c.$. 
By using the following relation derived from the generalized Ward identity,
\begin{align}
\label{Ward's identities}
(q/\hbar)\partial_{k_{\alpha}}G^{R/A}(\bm{k},\omega)=G^{R/A}(\bm{k},\omega)J_{\alpha}(\bm{k})G^{R/A}(\bm{k},\omega),
\end{align}
Eqs. \eqref{B5} and \eqref{B6} become
\begin{align}
A_0(\omega)&=\frac{q^2}{\hbar^2} \int \frac{d\bm{k}}{(2\pi)^d} f(\omega) \partial_{k_{\gamma}} \partial_{k_{\beta}} \mathrm{Tr} \bigl[ s_{\alpha}(G^R(\bm{k},\omega)-G^A(\bm{k},\omega))\bigr], \label{B11}\\
A_1(\omega)&=\frac{q}{\hbar} \int \frac{d\bm{k}}{(2\pi)^d}  \biggl\{ \biggl(\frac{\partial f(\omega)}{\partial \omega}\biggr) \partial_{k_\gamma} \mathrm{Tr} \bigl[ s_{\alpha} G^R(\bm{k},\omega)J_{\beta}(\bm{k}) G^A(\bm{k},\omega) \bigr] \notag \\
&\hspace{2.0cm} +f(\omega) \biggl( \partial_{k_\gamma}\mathrm{Tr} \biggl[ s_{\alpha} \frac{\partial G^R(\bm{k},\omega)}{\partial \omega} J_{\beta}(\bm{k}) G^R(\bm{k},\omega)\biggr] +c.c.\biggr) \biggr\}. \label{B12}
\end{align}
Eqs. \eqref{B11} and \eqref{B12} vanish because they can be written as an integration over a derivative.
Therefore, a divergence does not occur even if we take the DC limit, and $A_2(\omega)$ determines the current-induced NEE.

%% file: appendices/appendixC.tex
\section{BAND REPRESENTATION FOR THE DC LIMIT} \label{app:C}

In this section, we derive the band representation for the DC limit by taking the weak-scattering limit. 
We define the Green's function of the $n$-th band as 
\begin{align}
    G^{R/A}_n(\bm{k},\omega)=\frac{1}{\hbar\omega-\varepsilon_n(\bm{k})\pm i\eta},
\end{align}
where $\eta$ is the scattering rate.
Here, we shift the integral variable $\omega$ and include the chemical potential $\mu$ into the Fermi distribution function; $f(\omega)=(1+e^{\beta(\hbar \omega-\mu)})^{-1}$. 
Note that the weak scattering means that the scattering rate is sufficiently small compared to the kinetic energy of electrons, $1/\beta$,
the energy of incident photons, $\hbar \omega_1$, and the interband energy, $\varepsilon_n-\varepsilon_m$.


When we perform the partial integrals in Eq. \eqref{CNEE_GR}, we ignore the Fermi sea terms, and Eq. \eqref{CNEE_GR} becomes
\begin{align}
    \label{CNEE_partial}
    \zeta^{(2)}_{\alpha;\beta\gamma} 
    =&2 \hbar \sum_{n,m,l} \int \frac{d\bm{k}}{(2\pi)^d} \int^{\infty}_{-\infty} \frac{d \omega }{2\pi } \biggl(-\frac{\partial f(\omega)}{\partial \omega } \biggr)\mathrm{Im} \biggl( s^{nm}_{\alpha}(\bm{k}) \frac{\partial G_m^R(\bm{k},\omega)}{\partial \omega } J^{ml}_{\beta}(\bm{k})G_l^R(\bm{k},\omega) J^{ln}_{\gamma}(\bm{k})G_n^{R-A}(\bm{k},\omega) \notag \\
    &+\frac{1}{2} s^{nm}_{\alpha}(\bm{k}) \frac{\partial G_m^R(\bm{k},\omega)}{\partial \omega } J^{mn}_{\beta\gamma}(\bm{k})G_n^{R-A}(\bm{k},\omega) \biggr)+(\beta \leftrightarrow \gamma).
\end{align}
Furthermore, we divide $\zeta^{(2)}_{\alpha;\beta\gamma}$ as 
\begin{align}
    \zeta^{(2)}_{\alpha;\beta\gamma}=2\hbar^2 q^2 \int \frac{d\bm{k}}{(2\pi)^d} \Bigl(\zeta^{(2),A}_{\alpha;\beta\gamma}(\bm{k})+\zeta^{(2),B}_{\alpha;\beta\gamma}(\bm{k})+\zeta^{(2),C}_{\alpha;\beta\gamma}(\bm{k})+\zeta^{(2),D}_{\alpha;\beta\gamma}(\bm{k})+\zeta^{(2),E}_{\alpha;\beta\gamma}(\bm{k}) \Bigr) +(\beta \rightarrow \gamma),
\end{align}
where each component is given by 
\begin{align}
    \zeta^{(2),A}_{\alpha;\beta\gamma}(\bm{k})&=\frac{1}{2} \sum_n \int^{\infty}_{-\infty} \frac{d \omega}{2\pi}  \Bigl(-\frac{\partial f(\omega)}{\hbar \partial \omega}\Bigr)
    \mathrm{Im}\Bigl[ s_{\alpha}^{nn} \frac{\partial G^R_n(\omega)}{\partial \omega}v_{\beta \gamma }^{nn} G_n^{R-A}(\omega) \Bigr],  \label{CNEE_A} \\
    \zeta^{(2),B}_{\alpha;\beta\gamma}(\bm{k})&= \sum_n \int^{\infty}_{-\infty} \frac{d \omega}{2\pi}  \Bigl(-\frac{\partial f(\omega)}{\hbar \partial \omega}\Bigr)
    \mathrm{Im}\Bigl[ s_{\alpha}^{nn} \frac{\partial G^R_n(\omega)}{\partial \omega} v_\beta^{nn} G^R_n(\omega) v_\gamma^{nn} G^{R-A}_n(\omega) \Bigr], \label{CNEE_B} \\
    \zeta^{(2),C}_{\alpha;\beta\gamma}(\bm{k})&= \sum_{ \substack{n,m \\ m \neq n }} \int^{\infty}_{-\infty} \frac{d \omega}{2\pi}  \Bigl(-\frac{\partial f(\omega)}{\hbar \partial \omega}\Bigr)
    \mathrm{Im}\Bigl[ s_{\alpha}^{nn} \frac{\partial G^R_n(\omega)}{\partial \omega} v_\beta^{nm} G^R_m(\omega) v_\gamma^{mn} G^{R-A}_n(\omega) \Bigr], \label{CNEE_C}  \\
    \zeta^{(2),D}_{\alpha;\beta\gamma}(\bm{k})&=\sum_{ \substack{n,m \\ m \neq n }} \int^{\infty}_{-\infty} \frac{d \omega}{2\pi}  \Bigl(-\frac{\partial f(\omega)}{\hbar \partial \omega}\Bigr) 
    \mathrm{Im}\Bigl[ s_{\alpha}^{nm} \frac{\partial G^R_m(\omega)}{\partial \omega} v_\beta^{mn} G^R_n(\omega) v_\gamma^{nn} G^{R-A}_n(\omega) \Bigr],  \label{CNEE_D} \\
    \zeta^{(2),E}_{\alpha;\beta\gamma}(\bm{k})&=\frac{1}{2}\sum_{ \substack{n,m \\ m \neq n }} \int^{\infty}_{-\infty} \frac{d \omega}{2\pi}  \Bigl(-\frac{\partial f(\omega)}{\hbar \partial \omega}\Bigr)
    \mathrm{Im}\Bigl[ s_{\alpha}^{nm} \frac{\partial G^R_m(\omega)}{\partial \omega}v_{\beta \gamma }^{mn} G^{R-A}_n(\omega) \Bigr],  \notag \\
    &\hspace{1cm}+\sum_{\substack{n,m,l \\ m\neq n, l \neq n }} \int^{\infty}_{-\infty} \frac{d \omega}{2\pi}  \Bigl(-\frac{\partial f(\omega)}{\hbar \partial \omega}\Bigr) 
    \mathrm{Im}\ \Bigl[ s_{\alpha}^{nm} \frac{\partial G^R_m(\omega)}{\partial \omega} v_\beta^{ml} G^R_l(\omega) v_\gamma^{ln} G^{R-A}_n(\omega) \Bigr]. \label{CNEE_E}
\end{align}
where $v_{\alpha_1 \cdots \alpha_n}(\bm{k} )=q^{-n} J_{\alpha_1 \cdots \alpha_n}(\bm{k} )$ is a velocity operator. 
If we perform a partial integral on $\zeta^{(2),B}_{\alpha;\beta\gamma}(\bm{k})$, Eq. \eqref{CNEE_B} becomes
\begin{align}
    \zeta^{(2),B}_{\alpha;\beta\gamma}(\bm{k})= \frac{1}{2}\sum_n \int^{\infty}_{-\infty} \frac{d \omega}{2\pi}  \Bigl(\frac{\partial^2 f(\omega)}{\hbar \partial \omega^2}\Bigr)
    \mathrm{Im}\Bigl[ s_{\alpha}^{nn} G^R_n(\omega) v_\beta^{nn} G^R_n(\omega) v_\gamma^{nn} G^{R-A}_n(\omega) \Bigr]. \label{CNEE_B_partial} 
\end{align}
Assuming the weak-scattering limit yields 
\begin{align}
    G^{R-A}_n(\omega)=\frac{-2i \eta}{(\hbar \omega-\varepsilon_n)^2+\eta^2}\sim-\frac{2\pi i}{\hbar} \delta(\omega-\varepsilon_n/\hbar),  \label{GR-GA} 
\end{align}
where $\delta(x)$ is the Dirac delta function. 
Performing frequency integrals using this equation, we find
\begin{align}
    &\zeta^{(2),A}_{\alpha;\beta\gamma}(\bm{k})
    =-\frac{1}{2 \eta^2}  \sum_n \mathrm{Re} (s_{\alpha}^{nn} v_{\beta \gamma}^{nn}) \Bigl(-\frac{\partial f_n}{\partial \varepsilon_n}\Bigr), \label{CNEE_A_semi} \\
    &\zeta^{(2),B}_{\alpha;\beta\gamma}(\bm{k})
    =\frac{1}{2 \eta^2} \sum_n \mathrm{Re} ( s_{\alpha}^{nn} v_\beta^{nn}  v_\gamma^{nn})\Bigl(\frac{\partial^2 f_n}{\partial \varepsilon_n^2}\Bigr), \label{CNEE_B_semi} \\
    &\zeta^{(2),C}_{\alpha;\beta\gamma}(\bm{k})
    =-\frac{1}{ \eta^2} \sum_{ \substack{ n,m \\ m \neq n }} \mathrm{Im} \biggl[ i s_{\alpha}^{nn} \frac{ v_{\beta}^{nm} v_{\gamma}^{mn}}{\varepsilon_{nm}+i \eta} \biggr] \Bigl(-\frac{\partial f_n}{\partial \varepsilon_n}\Bigr), \label{CNEE_C_semi} \\
    &\zeta^{(2),D}_{\alpha;\beta\gamma}(\bm{k})
    =\frac{1}{\eta} \sum_{ \substack{ n,m \\ m \neq n }} \mathrm{Im} \biggl[  \frac{ s_{\alpha}^{nm} v_{\beta}^{mn}}{(\varepsilon_{nm}+i \eta)^2} v_{\gamma}^{nn} \biggr] \Bigl(-\frac{\partial f_n}{\partial \varepsilon_n}\Bigr), \label{CNEE_D_semi} \\
    &\zeta^{(2),E}_{\alpha;\beta\gamma}(\bm{k}) 
    =\frac{1}{2} \sum_{ \substack{ n,m \\ m \neq n }} \mathrm{Im} \biggl[i \frac{s_{\alpha}^{nm} v_{\beta \gamma }^{mn}}{(\varepsilon_{nm}+i \eta)^2} \biggr] \Bigl(-\frac{\partial f_n}{\partial \varepsilon_n}\Bigr) 
    +\sum_{\substack{n,m,l \\ m \neq n, l \neq n }} \mathrm{Im} \biggl[ i \frac{ s_{\alpha}^{nm} v_{\beta}^{ml} v_{\gamma}^{ln}}{(\varepsilon_{nm}+i \eta)^2 (\varepsilon_{nl}+i \eta)}  \biggr] \Bigl(-\frac{\partial f_n}{\partial \varepsilon_n}\Bigr),  \label{CNEE_E_semi}
\end{align}
where $\varepsilon_{nm}=\varepsilon_n-\varepsilon_m$ and $f_n=f(\varepsilon_n)=(1+e^{\beta (\varepsilon_n-\mu) })^{-1}$.
Furthermore, by expanding the powers of $\eta$ in a Taylor series up to the order $O(\eta)$,
Eqs. \eqref{CNEE_C_semi}$\sim$\eqref{CNEE_E_semi} can be rewritten as 
\begin{align}
    \zeta^{(2),C}_{\alpha;\beta\gamma}(\bm{k})  
    &=-\frac{1}{ \eta^2} \sum_{ \substack{ n,m \\ m \neq n }} \mathrm{Im} \biggl[ i s_{\alpha}^{nn} \frac{ v_{\beta}^{nm} v_{\gamma}^{mn}}{\varepsilon^2_{nm}+\eta^2} (\varepsilon_{nm}-i\eta) \biggr] \Bigl(-\frac{\partial f_n}{\partial \varepsilon_n}\Bigr)  \\
    &=-\frac{1}{ \eta^2} \sum_{ \substack{ n,m \\ m \neq n }} \mathrm{Im} \biggl[ i s_{\alpha}^{nn} \frac{ v_{\beta}^{nm} v_{\gamma}^{mn}}{\varepsilon_{nm}}\biggl( 1-\frac{\eta^2}{\varepsilon^2_{nm}}\biggr) 
    +\eta s^{nn}_\alpha \frac{ v_{\beta}^{nm} v_{\gamma}^{mn}}{\varepsilon^2_{nm}} \biggr] \Bigl(-\frac{\partial f_n}{\partial \varepsilon_n}\Bigr)+O(\eta) \\
    &= \sum_{ \substack{ n,m \\ m \neq n }} \biggl\{ -\frac{1}{ \eta^2} \mathrm{Re} \biggl[  s_{\alpha}^{nn} \frac{ v_{\beta}^{nm} v_{\gamma}^{mn}}{\varepsilon_{nm}} \biggr]+\mathrm{Re} \biggl[s_{\alpha}^{nn}\frac{ v_{\beta}^{nm} v_{\gamma}^{mn}}{\varepsilon^3_{nm}} \biggr] 
    -\frac{1}{\eta} \mathrm{Im}\biggl[ s_{\alpha}^{nn}\frac{ v_{\beta}^{nm} v_{\gamma}^{mn}}{\varepsilon^2_{nm}} \biggr] \biggr\}\Bigl(-\frac{\partial f_n}{\partial \varepsilon_n}\Bigr)+O(\eta), \\
    \zeta^{(2),D}_{\alpha;\beta\gamma}(\bm{k}) 
    &=\frac{1}{\eta} \sum_{ \substack{ n,m \\ m \neq n }} \mathrm{Im} \biggl[  \frac{ s_{\alpha}^{nm} v_{\beta}^{mn}}{(\varepsilon^2_{nm}+\eta^2)^2} v_{\gamma}^{nn}(\varepsilon_{nm}-i\eta)^2 \biggr] \Bigl(-\frac{\partial f_n}{\partial \varepsilon_n}\Bigr)  \\
    &= \sum_{ \substack{ n,m \\ m \neq n }} \biggl\{ \frac{1}{\eta} \mathrm{Im} \biggl[  \frac{ s_{\alpha}^{nm} v_{\beta}^{mn}}{\varepsilon_{nm}^2} v_{\gamma}^{nn} \biggr] 
    -2\mathrm{Re} \biggl[  \frac{ s_{\alpha}^{nm} v_{\beta}^{mn}}{\varepsilon_{nm}^3} v_{\gamma}^{nn} \biggr] \biggr\}\Bigl(-\frac{\partial f_n}{\partial \varepsilon_n}\Bigr)+O(\eta), \\
    \zeta^{(2),E}_{\alpha;\beta\gamma}(\bm{k}) 
    &=\frac{1}{2} \sum_{ \substack{ n,m \\ m \neq n }} \mathrm{Im} \biggl[i \frac{s_{\alpha}^{nm} v_{\beta \gamma }^{mn}}{(\varepsilon_{nm}^2+\eta^2)^2}(\varepsilon_{nm}-i\eta)^2 \biggr] \Bigl(-\frac{\partial f_n}{\partial \varepsilon_n}\Bigr) \notag \\
    &\hspace{0.4cm}+\sum_{\substack{n,m,l \\ m \neq n, l \neq n }} \mathrm{Im} \biggl[ i \frac{ s_{\alpha}^{nm} v_{\beta}^{ml} v_{\gamma}^{ln}}{(\varepsilon_{nm}^2+\eta^2)^2(\varepsilon_{nl}^2+\eta^2)} 
    (\varepsilon_{nm}-i\eta)^2(\varepsilon_{nl}-i\eta)  \biggr] \Bigl(-\frac{\partial f_n}{\partial \varepsilon_n}\Bigr)  \\
    &= \frac{1}{2} \sum_{ \substack{ n,m \\ m \neq n }}  \mathrm{Re}  \biggl[ \frac{s_{\alpha}^{nm} v_{\beta \gamma }^{mn}}{\varepsilon_{nm}^2} \biggr]\Bigl(-\frac{\partial f_n}{\partial \varepsilon_n}\Bigr) 
    +\sum_{\substack{n,m,l \\ m \neq n, l \neq n }} \mathrm{Re} \biggl[ \frac{ s_{\alpha}^{nm} v_{\beta}^{ml} v_{\gamma}^{ln}}{\varepsilon_{nm}^2 \varepsilon_{nl}} \biggr]\Bigl(-\frac{\partial f_n}{\partial \varepsilon_n}\Bigr)+O(\eta).
\end{align}
Replacing $\eta$ with $\hbar/\tau$ where $\tau$ is the relaxation time and classifying $\zeta^{(2)}_{\alpha;\beta\gamma}$ into the order of $\tau$, we can obtain
\begin{align}
    \zeta^{(2),\tau^2}_{\alpha;\beta\gamma} 
    &=q^2 \tau^2\sum_n \int \frac{d\bm{k}}{(2\pi)^d} \mathrm{Re} \biggl[ s_{\alpha}^{nn} v_\beta^{nn}  v_\gamma^{nn}\Bigl(\frac{\partial^2 f_n}{\partial \varepsilon_n^2}\Bigr) 
    +\Bigl(s_{\alpha}^{nn} v_{\beta \gamma}^{nn}+ 2 \sum_{m (\neq n)}s_{\alpha}^{nn} \frac{ v_{\beta}^{nm} v_{\gamma}^{mn}}{\varepsilon_{nm}} \Bigr)\Bigl(\frac{\partial f_n}{\partial \varepsilon_n} \Bigr) \biggr]+(\beta \rightarrow \gamma)  \\
    &=\frac{2q^2}{\hbar^2}\tau^2 \sum_n \int \frac{d\bm{k}}{(2\pi)^d} s^{nn}_\alpha \partial_{k_\beta} \partial_{k_\gamma} f_n, \label{CNEE_tau2_lim} \\
    \zeta^{(2),\tau}_{\alpha;\beta\gamma} 
    &=2 \hbar q^2 \tau \sum_{ \substack{ n,m \\ m \neq n }} \mathrm{Im} \biggl[  \frac{ s_{\alpha}^{nm} v_{\beta}^{mn}}{\varepsilon_{nm}^2} v_{\gamma}^{nn} \biggr] \Bigl(-\frac{\partial f_n}{\partial \varepsilon_n}\Bigr) 
    -\cancel{2 \hbar q^2 \tau \sum_{ \substack{ n,m \\ m \neq n }} \mathrm{Im} \biggl[ s^{nn}_\alpha \frac{ v_{\beta}^{nm} v_{\gamma}^{mn}}{\varepsilon_{nm}^2}  \biggr] \Bigl(-\frac{\partial f_n}{\partial \varepsilon_n}\Bigr)}+(\beta \rightarrow \gamma)   \label{CNEE_tau1_lim} \\
    &=- q^2 \tau \sum_{ \substack{ n,m \\ m \neq n }} 2 \mathrm{Im} \biggl[  \frac{ s_{\alpha}^{nm} v_{\beta}^{mn}}{\varepsilon_{nm}^2} \biggr] \partial_{k_\gamma}f_n+(\beta \rightarrow \gamma), \\
    \zeta^{(2),\tau^0}_{\alpha;\beta\gamma} 
    &= \hbar^2 q ^2 \sum_{ \substack{ n,m \\ m \neq n }} \int \frac{d\bm{k}}{(2\pi)^d} \mathrm{Re} \biggl[ \frac{s_{\alpha}^{nm}}{\varepsilon_{nm}^2} \biggl(v_{\beta \gamma }^{mn}+\sum_{l(\neq n)}2\frac{ v_{\beta}^{ml} v_{\gamma}^{ln}}{ \varepsilon_{nl}} \biggr) 
    +2s_{\alpha}^{nn}\frac{ v_{\beta}^{nm} v_{\gamma}^{mn}}{\varepsilon^3_{nm}}  -4\frac{ s_{\alpha}^{nm} v_{\beta}^{mn}}{\varepsilon_{nm}^3} v_{\gamma}^{nn} \biggr]\Bigl(-\frac{\partial f_n}{\partial \varepsilon_n}\Bigr)+(\beta \rightarrow \gamma). \notag \\ \label{CNEE_tau0_lim}
\end{align}
Here, we use 
\begin{align}
    \hbar^{-1} \partial_{k_\beta} v^{nn}_\gamma=v_{\beta \gamma}^{nn}+  \sum_{m (\neq n)} \frac{ v_{\beta}^{nm} v_{\gamma}^{mn}+v_{\gamma}^{nm}v_{\beta}^{mn}}{\varepsilon_{nm}},
\end{align}
to derive Eq. \eqref{CNEE_tau2_lim} and drop the second term of Eq. \eqref{CNEE_tau1_lim} because it cancels out with the term when the indices are interchanged.
The $\tau^2$ term and $\tau$ term are consistent with Eq. (68) in \cite{Fregoso2022-fd} and Eq. (7) in \cite{Xiao2023-yu}, respectively.
As for the $\tau^0$ term, the second and third terms in Eq. \eqref{CNEE_tau0_lim} agree with the Fermi surface terms of Eq. (9) in \cite{Xiao2022-xr}. 

%% file: appendices/appendixD.tex
\section{BAND REPRESENTATION FOR FINITE FREQUENCIES} \label{app:D}

In this section, we derive the band representation for finite frequencies by taking the weak-scattering limit. 
We first introduce the reduced density matrix (RDM) formalism in the velocity gauge and compare it with the result of taking the weak-scattering limit in Eq. \eqref{Second-order response function}.
Note that the RDM formalism in the length gauge is derived in Ref.~\cite{Fregoso2022-fd}.


\subsection{RDM formalism in the velocity gauge}
It is only necessary to consider the dynamics of the RDM $\hat{\rho}_{\bm{k}}(t)$ in the subspace $\mathbb{V}_{\bm{k}}$ labeled by the crystal momentum $\bm{k}$ instead of the full density matrix $\hat{\rho}(t)$ 
because we can express $\hat{\rho}(t)$ as the tensor product of the RDMs; $\hat{\rho}(t)=\prod_{\bm{k}} \otimes \hat{\rho}_{\bm{k}}(t)$.
The matrix representation of the RDM is
\begin{align}
    \rho_{\bm{k}nm}(t)=\mathrm{Tr}\bigl[ \hat{\rho}(t) \hat{c}_{\bm{k}m}^\dagger \hat{c}_{\bm{k}n}\bigr],
\end{align}
where $\hat{c}_{\bm{k}n}^\dagger$ and $\hat{c}_{\bm{k}n}$ are fermionic creation and annihilation operators.
According to the von Neumann equation $i\hbar \partial_t \hat{\rho}(t)=[\hat{H}(t),\hat{\rho}(t)]$, the equation of motion for the RDM can be described as
\begin{align}
    \label{eqm_RDM}
    i\hbar \partial_t \rho_{\bm{k}nm}(t)
    =\mathrm{Tr}\bigl[\hat{\rho}(t)[\hat{c}_{\bm{k}m}^\dagger \hat{c}_{\bm{k}n},\hat{H}(t)] \bigr].
\end{align}
We here assume $\hat{H}(t)=\hat{H}_0+\hat{V}(t)$ where $\hat{H}_0$ and $\hat{V}(t)$ are unperturbed and perturbed Hamiltonians by an external field $\bm{F}$, respectively which are described as  
\begin{align} 
    \hat{H}_0=\sum_n \int \frac{d \bm{k}}{(2\pi)^d} \varepsilon_{\bm{k}n} \hat{c}_{\bm{k}n}^\dagger \hat{c}_{\bm{k}n},
    \hspace{0.5cm} \hat{V}(t)=\sum_{n,m} \int \frac{d \bm{k}}{(2\pi)^d} \hat{c}_{\bm{k}n}^\dagger V_{\bm{k}nm}(t)\hat{c}_{\bm{k}m}.
\end{align}
By using the anticommutation relations
\begin{align}
   \{\hat{c}_{\bm{k}n}, \hat{c}_{\bm{k}'m}\}=\{\hat{c}_{\bm{k}n}^\dagger, \hat{c}_{\bm{k}'m}^\dagger\}=0, 
   \hspace{0.5cm}\{\hat{c}_{\bm{k}n}, \hat{c}_{\bm{k}'m}^\dagger\}=(2\pi)^d\delta_{nm}\delta(\bm{k}-\bm{k}'),
\end{align}
Eq. \eqref{eqm_RDM} becomes
\begin{align}
    \label{eqm_RDM_perturb}
    (i\hbar \partial_t-\varepsilon_{nm}) \rho_{\bm{k}nm}(t)=[V(t),\rho_{\bm{k}}(t)]_{nm},
\end{align}
where $[V(t),\rho_{\bm{k}}(t)]_{nm}=\sum_l (V_{nl}(t)\rho_{\bm{k}lm}(t)-\rho_{\bm{k}nl}(t)V_{lm}(t))$.
We here present a phenomenological treatment of the scattering rate \cite{Watanabe2022-qp}.
The scattering rate $\eta$ is introduced by modifying Eq. \eqref{eqm_RDM_perturb} to 
\begin{align}
    \label{eqm_RDM_scat}
    (i\hbar \partial_t-\varepsilon_{nm}) \rho_{\bm{k}nm}^{(p)}(t)=\sum_{q=0}^{p}[V^{(p-q)}(t),\rho_{\bm{k}}^{(q)}(t)]_{nm}-ip\eta \rho^{(p)}_{\bm{k}nm}(t),
\end{align}
where $\rho_{\bm{k}}^{(p)}$ and $V^{(p)}$ are $O(|\bm{F}|^p)$.
In other words, the scattering rate is multiplied by the perturbation order $p$ of the RDM.
Furthermore, taking the Fourier transformation results in  
\begin{align}
    \label{eqm_RDM_omega}
    (\hbar \omega -\varepsilon_{nm}+i p \eta) \rho^{(p)}_{\bm{k}nm}(\omega)= \int \frac{d\omega_1}{2\pi} \frac{d\omega_2}{2\pi }\sum_{q=0}^{p}[V^{(p-q)}(\omega_1),\rho_{\bm{k}}^{(q)}(\omega_2)]_{nm} 2\pi \delta_{\omega_1+\omega_2,\omega}. 
\end{align}
In the velocity gauge, we can describe $\hat{H}(t)$ as 
\begin{align}
    \hat{H}(t)&=\hat{H}_0\bigl( \bm{k}-\frac{q}{\hbar}\bm{A}(t)\bigr) 
    =\hat{H}_0(\bm{k})-q\hat{v}_\beta(\bm{k}) A^\beta(t)+\frac{q^2}{2} \hat{v}_{\beta \gamma}(\bm{k})A^\beta(t)A^\gamma(t)+\cdots. 
\end{align}
If we perfome the Fourier transformation, $V^{(p)}_{nm}(\omega)$ $(p=1,2)$ are given by
\begin{align}
    V^{(1)}_{nm}(\omega)&=-q v^{nm}_\beta \int \frac{d\omega_1}{2\pi}A^\beta(\omega_1) 2\pi \delta_{\omega_1,\omega},  \\
    V^{(2)}_{nm}(\omega)&=\frac{q^2}{2} v^{nm}_{\beta \gamma} \int \frac{d\omega_1}{2\pi} \frac{d\omega_2}{2\pi}A^\beta(\omega_1)A^\gamma(\omega_2) 2\pi \delta_{\omega_1+\omega_2,\omega}. 
\end{align}
Therefore, the $p$-th order RDMs $(p=0,1,2)$ become $\rho_{\bm{k}nm}^{(0)}(\omega) \equiv\delta_{nm}f_m 2\pi\delta(\omega)$ and 
\begin{align}
    &\rho^{(1)}_{\bm{k}nm}(\omega) = \int \frac{d\omega_1}{2\pi} \frac{q v_\beta^{nm} f_{nm}}{\hbar \omega_1-\varepsilon_{nm}+i\eta}A^\beta(\omega_1)2\pi \delta_{\omega_1,\omega}, \label{rho1} \\
    &\rho^{(2)}_{\bm{k}nm}(\omega) = \int \frac{d\omega_1}{2\pi} \frac{d\omega_2}{2\pi } A^\beta(\omega_1)A^\gamma(\omega_2)2\pi \delta_{\omega_1+\omega_2,\omega} \notag \\
    &\hspace{1.54cm}\times q^2\biggl\{ \frac{1}{2} \frac{ v^{nm}_{\beta \gamma} f_{mn}}{\hbar \omega-\varepsilon_{nm}+2i\eta}+\sum_l\frac{1}{\hbar \omega-\varepsilon_{nm}+2i\eta} 
    \biggl( \frac{ v_\beta^{nl}v _\gamma^{lm} f_{ml}}{\hbar \omega_2-\varepsilon_{lm}+i\eta}- \frac{ v_\gamma^{nl}v _\beta^{lm} f_{ln}}{\hbar \omega_2-\varepsilon_{nl}+i\eta}\biggr)\biggr\}, \label{rho2}
\end{align}
where $f_{nm}=f_n-f_m$. The expectation value of the spin density is calculated by the spin operator and the RDM as follows;
\begin{align}
    \label{spin_density_w}
    \braket{\delta\hat{s}_\alpha (\omega)}=\sum_{n,m} \int \frac{d \bm{k}}{(2\pi)^d} s^{nm}_\alpha (\bm{k}) \rho_{\bm{k} mn} (\omega).
\end{align}
In particular, the second-order spin density $\braket{\delta\hat{s}_\alpha (\omega)}^{(2)}$ is 
\begin{align}
    \braket{\delta\hat{s}_\alpha (\omega)}^{(2)} 
    &=2\sum_{n,m} \int \frac{d \bm{k}}{(2\pi)^d} s^{nm}_\alpha \rho^{(2)}_{\bm{k}mn}(\omega) \label{RDM_2nd_velocity_1}  \\
    &=\int \frac{d\omega_1}{2\pi} \frac{d\omega_2}{2\pi}E^\beta(\omega_1)E^\gamma(\omega_2)2\pi\delta_{\omega_1+\omega_2,\omega} \notag \\
    &\times-\frac{q^2}{(\omega_1+i\delta)(\omega_2+i\delta)} \sum_{n,m} \int \frac{d\bm{k}}{(2\pi)^d}\biggl\{ \frac{1}{2} \frac{ s^{nm}_\alpha v^{mn}_{\beta \gamma}f_{nm}}{\hbar \omega-\varepsilon_{mn}+2i\eta} \notag \\
    &\hspace{2cm}+\sum_l\frac{s_\alpha^{nm}}{\hbar \omega-\varepsilon_{mn}+2i\eta} \biggl( \frac{ v_\beta^{ml}v _\gamma^{ln} f_{nl}}{\hbar \omega_2-\varepsilon_{ln}+i\eta}- \frac{  v_\gamma^{ml}v _\beta^{ln} f_{lm}}{\hbar \omega_2-\varepsilon_{ml}+i\eta}\biggr)\biggr\}+\bigl[(\beta,\omega_1)\leftrightarrow(\gamma,\omega_2)\bigr]. \label{RDM_2nd_velocity_2}
\end{align}
Here, we multiply the factor 2 in the first line of the above expressions to include the term interchanging the indices and frequencies of the electric fields.
This additional contribution is reflected in the term, $\bigl[(\beta,\omega_1)\leftrightarrow(\gamma,\omega_2)\bigr]$.
Furthermore, we replace the vector potential $A$ with the electric field $E$ by using Eq. \eqref{AtoE}.


\subsection{Weak scattering limit in Eq. \eqref{Second-order response function}}

If we consider Eq. \eqref{GR-GA}, we can describe Eq. \eqref{Second-order response function} as
\begin{align}
    \zeta^{(2)}_{\alpha;\beta \gamma}(\omega;\omega_1,\omega_2) 
    &=-\frac{q^2}{(\omega_1+i\delta)(\omega_2+i\delta)} \sum_{n,m}\int \frac{d \bm{k}}{(2\pi)^d} \biggl\{ \frac{1}{2} \frac{s^{nm}_\alpha v^{mn}_{\beta \gamma} f_{nm}}{\hbar \omega-\varepsilon_{mn}+i\eta} \notag \\
    &\hspace{0.4cm}+\sum_l \frac{s^{nm}_\alpha v^{ml}_\beta v^{ln}_\gamma}{(\hbar \omega-\varepsilon_{mn}+i\eta)(\hbar \omega_1-\varepsilon_{ml}+i\eta)(\hbar\omega_2-\varepsilon_{ln}+i\eta)} \notag \\
    &\hspace{0.4cm}\times \Bigl( (\hbar \omega_1-\varepsilon_{ml}+i\eta)f_n-(\hbar \omega -\varepsilon_{mn}+i\eta)f_l 
    +(\hbar \omega_2-\varepsilon_{ln}+i\eta)f_m \Bigr)\biggr\}+\bigl[(\beta,\omega_1)\leftrightarrow(\gamma,\omega_2)\bigr],
\end{align}
where we change $\omega_\Sigma$ with $\omega$.
Here, we replace $\hbar \omega+i\eta$ with $\hbar \omega+2i\eta$ for a technical reason. 
We note that this modification changes the results in the region of the peak ($\hbar \omega=\varepsilon_{nm}$) and in the low-frequency region for the diagonal response ($n=m$), as already pointed out in Refs.~\cite{Parker2019-pa,Passos2018-tl}.
Using this modification, we find
\begin{align}
    \zeta^{(2)}_{\alpha;\beta \gamma}(\omega;\omega_1,\omega_2)&=-\frac{q^2}{(\omega_1+i\delta)(\omega_2+i\delta)} \sum_{n,m}\int \frac{d \bm{k}}{(2\pi)^d} \biggl\{ \frac{1}{2} \frac{s^{nm}_\alpha v^{mn}_{\beta \gamma} f_{nm}}{\hbar \omega-\varepsilon_{mn}+2i\eta} \notag \\
    &\hspace{0.4cm}+\sum_l \frac{s^{nm}_\alpha }{\hbar \omega-\varepsilon_{mn}+2i\eta} \biggl( \frac{v^{ml}_\beta v^{ln}_\gamma f_{nl}}{\hbar\omega_2-\varepsilon_{ln}+i\eta}-\frac{v^{ml}_\gamma v^{ln}_\beta f_{lm}}{\hbar \omega_2-\varepsilon_{ml}+i\eta}\biggr)\biggr\}+\bigl[(\beta,\omega_1)\leftrightarrow(\gamma,\omega_2)\bigr],
\end{align}
which is consistent with Eq. \eqref{RDM_2nd_velocity_2}.
\end{widetext}